\documentclass[12pt,fleqn]{article}

\usepackage{amsmath,amssymb,graphicx}

\usepackage[textwidth=15.9cm,textheight=23.2cm]{geometry}
\usepackage{latexsym}
\usepackage{cite}
\usepackage{bbm}
\usepackage{mathrsfs}

\numberwithin{equation}{section}

\def\AdSs5{$AdS_5$}
\def\AdSS5{$AdS_5$}
\def\AdS5s5{$AdS_5 \times S^5$}
\def\al{{\alpha^{\prime}}}
\def\gs{g_{st}}
\def\gy{g_{_{\rm YM}}}
\def\er{{\rm e}}
\def\dr{{\rm d}}

\def\tr{{\rm tr}}
\def\gs{g_{\rm s}}

\newcommand{\RR}{${\rm R}\!\otimes\!{\rm R}$\ }

\newcommand{\eg}{{\it e.g.~}}
\newcommand{\ie}{{\it i.e.~}}

\newcommand{\del}{\partial}

\newcommand{\be}{\begin{equation}}
\newcommand{\ee}{\end{equation}}
\newcommand{\ba}{\begin{eqnarray}}
\newcommand{\ea}{\end{eqnarray}}
\newcommand{\bdm}{\begin{displaymath}}
\newcommand{\edm}{\end{displaymath}}
\newcommand{\ra}{\rangle}
\newcommand{\la}{\langle}
\newcommand{\pp}{\prime}

\newcommand\fr[1]{\frac{1}{#1}}
\newbox\SlashedBox
\def\fs#1{\setbox\SlashedBox=\hbox{#1}
\hbox to
0pt{\hbox to 1\wd\SlashedBox{\hfil/\hfil}\hss}{#1}}
\def\hboxtosizeof#1#2{\setbox\SlashedBox=\hbox{#1}
\hbox to
1\wd\SlashedBox{#2}}

\def\ms#1{\setbox\SlashedBox=\hbox{$#1$}
\hbox to 0pt{\hbox to
1\wd\SlashedBox{\hfil/\hfil}\hss}#1}

\def\t2{\tau_2}
\def\IZ{\relax\ifmmode\mathchoice {\hbox{\cmss Z\kern-.4em Z}}
{\hbox{\cmss Z\kern-.4em Z}}
{\lower.9pt\hbox{\cmsss Z\kern-.4em Z}}
{\lower1.2pt\hbox{\cmsss Z\kern-.4em Z}}
\else{\cmss Z\kern-.4em Z}\fi}

\def\b{\beta}
\def\a{{\alpha}}
\def\g{\gamma}
\def\veps{\varepsilon}

\def\adot{{\dot\alpha}}

\def\d{\delta}

\def\c1{{\chi^1}}

\def\v{\varphi}

\def\N4{{\cal N}=4}
\def\half{\frac{1}{2}}

\def\nn{\nonumber}

\newcommand{\ups}{\upsilon}
\newcommand{\calA}{{\mathcal A}}

\newcommand{\calF}{{\mathcal F}}

\newcommand{\calH}{{\mathcal H}}

\newcommand{\calL}{{\mathcal L}}
\newcommand{\calM}{{\mathcal M}}
\newcommand{\calN}{{\mathcal N}}

\newcommand{\calQ}{{\mathcal Q}}
\newcommand{\calR}{{\mathcal R}}

\newcommand{\calW}{{\mathcal W}}

\newcommand{\scrH}{{\mathscr H}}

\newcommand{\scrN}{{\mathscr N}}

\newcommand{\scrQ}{{\mathscr Q}}

\newcommand{\scrS}{{\mathscr S}}

\newcommand{\scrV}{{\mathscr V}}
\newcommand{\scrW}{{\mathscr W}}

\DeclareMathAlphabet{\mathpzc}{OT1}{pzc}{m}{it}
\newcommand\hsp[1]{\hspace*{#1 cm}}
\newcommand\vsp[1]{\vspace*{#1 cm}}
\newcommand{\ndt}{\noindent}
\newcommand\atmp[3]{{\it Adv.\ Theor.\ Math.\ Phys.\ }{\bf #1} (#2) #3}

\newcommand\cqg[3]{{\it Class.\ and Quant.\ Grav.\ }{\bf #1} (#2) #3}

\newcommand\epjc[3]{{\it Eur.\ Phys.\ J. }{\bf C#1} (#2) #3}

\newcommand\ijtp[3]{{\it Int.\ J.\ Theor.\ Phys.\ }{\bf #1} (#2) #3}

\newcommand\jgp[3]{{\it J.\ Geom.\ Phys.\ }{\bf #1} (#2) #3}
\newcommand\jhep[3]{{\it J. High Energy Phys.\ }{\bf #1} (#2) #3}

\newcommand\npb[3]{{\it Nucl.\ Phys.\ }{\bf B#1} (#2) #3}

\newcommand\nc[3]{{\it Nuovo Cim.\ }{\bf #1} (#2) #3}

\newcommand\pla[3]{{\it Phys.\ Lett.\ }{\bf A#1} (#2) #3}
\newcommand\plb[3]{{\it Phys.\ Lett.\ }{\bf B#1} (#2) #3}

\newcommand\prd[3]{{\it Phys.\ Rev.\ }{\bf D#1} (#2) #3}

\newcommand\prep[3]{{\it Phys.\ Rept.\ }{\bf #1} (#2) #3}
\newcommand\prl[3]{{\it Phys.\ Rev.\ Lett.\ }{\bf #1} (#2) #3}

\newcommand\rmp[3]{{\it Rev.\ Mod.\ Phys.\ }{\bf #1} (#2) #3}

\newcommand{\hepth}[1]{{\tt hep-th/#1}}

\newcommand{\mb}[1]{\mathbf{#1}}

\newcommand{\mbit}[1]{\mbox{\textit{\textbf{#1}}}}
\begin{document}


\thispagestyle{empty}
\renewcommand{\thefootnote}{\fnsymbol{footnote}}

{\hfill \parbox{4cm}{
        HU-EP-04/09\\
        AEI-2004-21 \\
}}

\bigskip\bigskip

\begin{center} \noindent \Large \bf
Holography and the Higgs branch of $\calN$=2 SYM theories
\end{center}

\bigskip\bigskip\bigskip

\centerline{Zachary Guralnik$^{a}$, Stefano Kovacs$^b$ 
and Bogdan Kulik$^b$
\footnote[1]{ zack@physik.hu-berlin.de, stefano.kovacs@aei.mpg.de,
bogdan.kulik@aei.mpg.de}}
\bigskip
\bigskip\bigskip

\centerline{$^a$ \it Institut f\"ur Physik} \centerline{\it
Humboldt-Universit\"at zu Berlin} \centerline{\it Newtonstra{\ss}e
15} \centerline{\it 12489 Berlin, Germany}
\bigskip
\centerline{$^b$ \it Max-Planck-Institut f\"ur Gravitationsphysik}
\centerline{\it Albert-Einstein-Institut} \centerline{\it Am
M\"uhlenberg 1, D-14476 Golm, Germany}
\bigskip\bigskip

\bigskip\bigskip

\renewcommand{\thefootnote}{\arabic{footnote}}

\centerline{\bf \small Abstract}
\medskip

{\small  
\ndt 
We present a proposal for the description of the Higgs
branch of four-dimensional $\calN$=2 supersymmetric Yang--Mills
theories in the context of the AdS/CFT correspondence. We focus on a
finite Sp($N$) $\calN$=2 theory arising as dual of a configuration of
$N$ D3-branes in the vicinity of four D7-branes and an orientifold
7-plane in type I$^\pp$ string theory. The field theory contains
hypermultiplets in the second rank anti-symmetric and in the
fundamental representations.  The Higgs branch has a dual description
in terms of gauge field configurations with non-zero instanton number
on the world-volume of the D7-branes. In this setting the
non-renormalisation of the metric on the Higgs branch implies
constraints on the $\al$ corrections to the D7-brane effective action,
including couplings to the curvature and five-form field strength.  In
the second part of the paper we discuss non-renormalisation properties
of BPS Wilson lines, which are closely related to the physics of the
Higgs branch. Using a formulation of the four-dimensional $\calN$=2
theory in terms of a three-dimensional $\calN$=2 superspace we show
that the expectation value of certain Wilson-line operators with
hypermultiplets at the end points is independent of the length and
thus coincides with the expectation value of the local operators
parametrising the Higgs branch.}

\newpage

\setcounter{page}{1}

\section{Introduction}
\label{intro}

The holographic string/gauge theory equivalence known as AdS/CFT
duality \cite{Maldacena:1997re,Gubser:1998bc,Witten:1998qj} has
provided new insights into numerous aspects of both string theory and
strongly coupled gauge theory, and great effort has gone into building
the AdS/CFT dictionary. In this article we formulate a proposal for
the description of the Higgs branch of $\calN$=2 supersymmetric
Yang--Mills (SYM) theories. This will allow us to obtain an AdS
realization of a well known non-renormalization theorem in four
dimensional ${\cal N} =2$ theories, namely the non-renormalization of
the metric on the Higgs branch. Besides adding this entry to the
AdS/CFT dictionary, we shall discuss how, reversing the point of view,
the non-renormalisation theorems known in field theory can be used in
this set-up to obtain constraints on the non-abelian
Dirac-Born-Infeld (DBI) action in a curved background with
Ramond-Ramond flux.

To discuss the AdS description of the Higgs branch, we will focus on a
${\mathcal N} =2$ Sp($N$) gauge theory~\footnote{In our conventions
the algebra sp($N$) corresponds to $C_N$, \ie we define Sp($N$) so
that Sp(1)=SU(2).} dual to a $Z_2$ orientifold of AdS$_5 \times S^5$
\cite{Fayyazuddin:1998fb,Aharony:1998xz}. This background is the near
horizon limit of $N$ D3-branes which are coincident with a tadpole
cancelling configuration of D7-branes and a negative charge
O7-plane. In the near horizon limit on the D3-branes the O7-plane and
D7-branes fill AdS$_5$ and wrap the $S^3$ inside $S^5$, which is a
fixed surface of the orientifold. The field content of the dual
$\calN$=2 theory consists of a vector multiplet, one hypermultiplet in
the second rank antisymmetric representation and four hypermultiplets
in the fundamental representation, resulting in a vanishing
$\b$-function. In the limit of vanishing expectation values for the
scalars the theory is believed to be exactly conformal. We shall
instead construct an AdS description of the Higgs branch of this
model. On the field theory side the Higgs branch corresponds to vacua
with non-zero expectation values for the scalars in the fundamental
hypermultiplets. The bulk dual description involves turning on a SO(8)
field strength with non-zero instanton number in the D7-brane world
volume. Such instantons can in turn be viewed as D3-branes dissolved
into the world volume of the D7-branes.

The world-volume effective action of the D7-branes is a non-abelian
gauge theory whose form in flat space is known only to low orders in
an expansion in $\alpha'$
\cite{Gross:1986iv,Tseytlin:1986ti,Tseytlin:1997cs,Bergshoeff:2001dc,
Koerber:2001uu,Koerber:2002zb}. The leading term in the D7-brane
action is quadratic in field strengths but, due to the embedding in AdS$_5
\times$ S$^5/Z_2$, differs from flat space
Yang--Mills theory. Nevertheless, we shall find that this term admits
solutions corresponding to field strengths which are anti-self-dual
with respect to a {\it flat} metric ($F^+ =0$), namely the usual
instanton solutions in ${\mathbb R}^4$.

For the ${\cal N}=2$ theory which we consider there is an exact
equivalence between the the Higgs branch and the moduli space of
Yang--Mills instantons
\cite{Douglas:1995bn,Witten:1995gx,Douglas:1996uz} (see
\cite{gk,Dorey:2002ik} for a review).  Realizing this correspondence
in the dual AdS description requires that instanton configurations be
exact solutions of the full D7-brane  action in the AdS background,
including $\al$ corrections.  Note that, using the AdS/CFT dictionary,
the $\alpha'$ expansion of the  D7-brane effective action is converted
into an expansion in $1/\sqrt{\lambda}$, where $\lambda$ is the 't
Hooft coupling of the dual four-dimensional ${\cal N}=2$ super
Yang--Mills theory.

The above requirement implies independent conditions that must be satisfied
separately by terms in the effective action involving only gauge
fields or terms which involve couplings to bulk fields with 
non-zero background value,
namely the curvature and \RR five-form. 
We shall first consider terms of the first type, involving only powers
of the gauge field strength. There are no $F^3$ terms at order $\al$
and thus the leading corrections come from $F^4$ couplings that arise
at order $\al^2$. We will show that taking such terms into account a
self-dual field strength, $F^+=0$, remains a solution of the
field equations. 

While the $F^4$ terms are known exactly (at least in flat space)
\cite{Gross:1986iv,Tseytlin:1986ti,Tseytlin:1997cs,Bergshoeff:2001dc},
little is known about other relevant terms that might modify the field
equations for the gauge potential on the D7-brane world-volume. These
include couplings to the curvature, \RR five-form and their
derivatives which are non-vanishing in the AdS background we
consider. Among the terms that need to be considered are those of
the form $D^r\calR^mD^s\calF_5^nF^2$ arising at order
$(\al)^{m+\half(n+r+s)}$. Here $\calR$ generically denotes the
curvature and $\calF_5$ the self-dual \RR five-form. In order for the
self-dual field configurations to remain solutions of the complete
field equations the sum of these terms 
must vanish in the AdS background.  This constraint holds at each order 
in $\alpha'$.

Some of the terms of order $\al^2$ involving two powers of $\calR$ can
be deduced from flat-space calculations of two-graviton disk
amplitudes \cite{Bachas:1999um,Fotopoulos:2001pt,Wijnholt:2003pw}. 
We shall verify that these known terms do not vanish in the AdS
background and therefore we shall argue that other terms must be present
in the D7-brane effective action at the same order. These may include
couplings to the \RR five-form, but also couplings to pull-backs
of the bulk Ricci tensor which are not fixed by the disk amplitudes
computed in a background with vanishing Ramond-Ramond flux. 

In addition to preserving the instanton solutions, the D7-brane
action in the AdS background must give the correct metric on the
Higgs branch. The hyper-K\"{a}hler metric on the Higgs branch of
four-dimensional ${\cal N}=2$ gauge theories is exactly given by
the tree level result \cite{Argyres:1996eh}. For the particular
${\cal N}=2$ theory we consider, this metric is also given by the
moduli space approximation for the dynamics of slowly varying
instantons in eight-dimensional super Yang--Mills theory
\cite{Douglas:1995bn,Witten:1995gx,Douglas:1996uz,Dorey:2002ik}.
The dynamics of slowly varying instantons on a D7-brane wrapping
AdS$_5 \times S^3$ must be described by the same metric. We will
find that the leading term in the $\alpha'$ (or
$1/\sqrt{\lambda}$) expansion of the D7 action gives the exact
metric on the Higgs branch. All the sub-leading terms must therefore 
give a vanishing contribution. We will verify that this is the case to
order $1/\lambda$, assuming the sum of the five-form and
curvature couplings to $F^2$ vanish in the AdS background.

Although we have not analysed this aspect here, using our proposal for
the description of the Higgs branch in the AdS/CFT context it should
be possible to compute correlation functions of composite operators in
the gauge theory utilising the same prescription as in the conformal
phase. Correlation functions should be obtained from supergravity
amplitudes involving appropriately modified bulk-to-boundary
propagators, which encode the information about the non-trivial field
strength on the world volume of the D7-branes.

In the second half of this article we shall discuss
non-renormalization properties for a class of Wilson line operators in
theories with eight supercharges. This part of our discussion is a
sequel to \cite{Guralnik:2003di}, in which a non-renormalization
theorem for certain BPS Wilson loops in maximally supersymmetric
Yang--Mills theories was proven using field theoretic methods. The
eight supercharge variant of this non-renormalization theorem is
closely related to the physics of the Higgs branch. We
will find that the vacuum expectation value (vev) of certain straight
BPS Wilson lines with hypermultiplet fields at the end points is
independent of the length and thus coincides with the expectation
value of local operators. These vev's of bilinear local operators
made of scalar components of the hypermultiplets are the quantities
which parametrise the Higgs branch.

In the case of four-dimensional ${\cal N}=2$ SYM theories, we will
obtain this result by using a three-dimensional ${\cal N} =2$
superspace which makes a very useful subgroup of the four-dimensional
${\cal N} =2$ supersymmetry manifest. Our result then follows from
equations satisfied by the ${\cal N}=2$, $d=3$ chiral ring. The same
results apply to other eight-supercharge theories obtained by
dimensional reduction. In five dimensions, the result is modified by a
generalized Konishi anomaly which is crucial for the validity of
proposals relating effective superpotentials to bosonic matrix models
\cite{Dijkgraaf:2002fc,Dijkgraaf:2002dh,Dijkgraaf:2003xk,
Cachazo:2003yc,Bena:2003tf}.

The organization of this paper is as follows. In section 2, we review
the AdS construction of an ${\cal N}=2$ theory with fundamental
hypermultiplets. In section 3, we show how to construct the AdS
description of the Higgs branch, working to leading order in
$\alpha'$. In section 4 we discuss the $\alpha'$ corrections and the
constraints imposed on curvature and five-form couplings by the
correspondence between the Higgs branch and Yang--Mills instantons. In
section 5, we verify to order $\alpha'^2$ that the metric on the
Higgs branch is correctly given by the dynamics of slowly moving
instantons on a D7-brane wrapping AdS$_5 \times S^3$, assuming that
couplings of the curvature and five-form to $F^2$ sum to zero in the
AdS background. In section 6, we present a new non-renormalization
theorem for a class of straight open Wilson lines in eight-supercharge
SYM theories.

\section{AdS/CFT duality for ${\cal N}=2$ theories with fundamental
representations}
\label{dualit}

We will consider a finite ${\cal N}=2$ gauge theory with gauge group
Sp($N$) and matter content comprising one hypermultiplet in the second
rank anti-symmetric representation~\footnote{The anti-symmetric
representation of Sp($N$) is reducible, containing the singlet plus a
$N(2N-1)-1$ dimensional irreducible representation. Throughout this
paper we will refer to the latter as the second rank anti-symmetric
representation.} and four hypermultiplets in the fundamental
representation. The conformal phase of this theory was first studied
in the AdS/CFT context in \cite{Fayyazuddin:1998fb,Aharony:1998xz} at
the perturbative level. The correspondence was subsequently
generalised to include instanton effects in \cite{mg,gns,th} and the
Penrose limit for this system was studied in \cite{bgmnn}. The AdS
dual is obtained from the type IIB orientifold $T^2/(-1)^F\Omega I$,
where $\Omega$ is the world sheet parity and $I$ the inversion,
$z\to-z$, on the torus $T^2$ \cite{sen}. At each of the four fixed
points of $I$ there are an orientifold 7-plane and four D7-branes. The
resulting type I$^\pp$ theory has SO(8)$^4$ gauge symmetry, with one
SO(8) factor associated with each group of D7-branes. Considering a
stack of $N$ D3-branes probing the region near one of the O7-planes
and taking the near horizon limit leads to the AdS$_5\times S^5/Z_2$
geometry with four D7-branes filling AdS$_5$ and wrapping the $S^3$
inside $S^5$ which is fixed under the orbifold action. The gravity
dual of the Sp($N$) gauge theory is therefore a theory of closed and
open strings. In the duality closed string states correspond to gauge
invariant composite operators written as traces of fields in the
adjoint  representation, whereas open strings propagating in the bulk
are associated with meson-like operators containing fields in the
fundamental and anti-symmetric representations. The isometries and
SO(8) gauge symmetry of the string theory correspond to conformal and
global symmetries  in the gauge theory.

The near horizon geometry on the D3-branes is AdS$_5 \times S^5/Z_2$,
with metric
\be 
\dr s^2 = \frac{r^2}{L^2} \left(-\dr x_0^2 + 
\dr x_1^2+\dr x_2^2+\dr x_3^2 \right) + \frac{L^2}{r^2} \left(\dr r^2 
+ r^2 \dr\hat\Omega_5^2 \right) \, ,
\label{fullD3}
\ee
where $L$ is the radius of both the AdS$_5$ and the $S^5$ factors and
as usual $L^4=4\pi\gs N\al^2$. In (\ref{fullD3}) we have denoted with
$x_\mu$, $\mu=0,1,2,3$, the coordinates on the AdS$_5$ boundary and
with $r$ the radial coordinate transverse to the D3-branes, $r^2 =
X_4^2 + \cdots +X_9^2$.  In (\ref{fullD3}) $\dr\hat\Omega_5^2$ denotes
the metric on $S^5/Z_2$ given by
\be
\dr\hat\Omega_5^2 = \dr\theta^2 + \sin^2\theta\,\dr\phi^2 +
\cos^2\theta \, \dr\Omega_3^2 \, ,
\label{S5Z2metric} 
\ee
where the range of $\phi$ is $[0,\pi]$ instead of $[0,2\pi]$ as for an
ordinary $S^5$. 

The D7-branes are at a fixed point of the orientifold,
$X_8=X_9=0$. After taking the near horizon limit they fill AdS$_5$ and
wrap the $S^3$ corresponding to $\theta=0$ in (\ref{S5Z2metric}),
which is fixed under $Z_2$. The induced metric on the D7-branes is
\be
\dr s^2 = \frac{U^2}{L^2} \, 
\dr x_\parallel^2 + \frac{L^2}{U^2}\left(\dr U^2 + U^2 \dr\Omega_3^2 
\right) = \ups^2 \, \dr x_\parallel^2 + \fr{\ups^2}\,\dr X_\perp^2 \, ,
\label{mmetric}
\ee
where
\bdm
U^2 = r^2\big|_{X_8=X_9=0} = X_4^2 + X_5^2 +X_6^2 + X_7^2 
\edm
 and 
\bdm 
\dr x_\parallel^2 = -\dr x_0^2 + \dr x_1^2 + \dr x_2^2 +\dr x_3^2 \, , 
\qquad \dr X_\perp^2 = \dr X_4^2 + \dr X_5^2 +\dr X_6^2 + \dr X_7^2 \, .
\edm
For convenience of notation in (\ref{mmetric}) we have also defined the
dimensionless variable $\ups$ related to $U$ by $\ups^2=U^2/L^2$. 

The isometry group of the background is thus
SO(2,4)$\times$SO(4)$\times$SO(2), where the SO(2,4) factor
corresponds to the isometries of AdS$_5$ and the remaining two factors
to rotations in the $S^3$ directions and in the $X^8$ and $X^9$
directions transverse to the D7-branes. Moreover there is a SO(8)
gauge symmetry associated with the open strings on the D7-branes.  In
the dual field theory these become conformal and global symmetries and
it will be convenient to rewrite them as
SO(2,4)$\times$SU(2)$_L\times$SU(2)$_R\times$U(1)$_R\times$SO(8)  in
order to classify the various fields according to their
transformation.  Here SO(2,4) is the conformal group of the
four-dimensional theory, SU(2)$_R\times$U(1)$_R$ is the $\calN$=2
R-symmetry and SU(2)$_L$ together with SO(8) form a global `flavour'
symmetry. The fields in the $\calN$=2 vector multiplet transforming in
the adjoint of Sp($N$), that we denote by
$(A_\mu,\v,\lambda_\a,\bar\lambda_\a)$, are singlets of SO(8) and
SU(2)$_L$. The gauge field $A_\mu$ is also a singlet of SU(2)$_R$ and
is not charged under U(1)$_R$.  The complex scalars, $\v$ and
$\v^\dagger$, are SU(2)$_R$ singlets and have U(1)$_R$ charge $\pm
2$. The fermions, $\lambda_\a$ and $\bar\lambda_\adot$, transform in
the $\mb2$ of SU(2)$_R$ and have U(1)$_R$ charge $\pm 1$. The
hypermultiplet in the second rank anti-symmetric tensor representation
of Sp($N$) contains scalars, $\phi$ and $\widetilde\phi$, in the
$(\mb2,\mb2)$ of SU(2)$_R\times$SU(2)$_L$ with zero U(1)$_R$ charge
and fermions, $(\psi_\a,\widetilde\psi_\a)$ and
$(\bar\psi_\adot,\,\widetilde{\!\bar\psi}_\adot)$, which are singlets
of SU(2)$_R$, transform in the $\mb2$ of SU(2)$_L$ and have U(1)$_R$
charge $\mp 1$ respectively. All these fields are also SO(8)
singlets. Finally the hypermultiplets in the  fundamental of Sp($N$)
contain complex scalars, $q$ and $\widetilde q$, which are in the
$\mb2$ of SU(2)$_R$ with zero U(1)$_R$ charge and are singlets of
SU(2)$_L$ and fermions, $(\eta_\a,\widetilde\eta_\a)$ and
$(\bar\eta_\adot,\widetilde{\bar\eta}_\adot)$, which are singlets of
SU(2)$_L\times$SU(2)$_R$ and have U(1)$_R$ charge $\pm 1$. The fields
in the fundamental hypermultiplets transform in the $\mb8_{\rm v}$ of
SO(8). The SU(2)$_L\times$SU(2)$_R\times$U(1)$_R\times$SO(8) quantum
numbers of the elementary fields are summarised in table \ref{tqn}.
\begin{table}
\begin{center}
\begin{tabular}{|c|c|c|c|c|c|c|}
\cline{3-7}
\multicolumn{2}{c|}{\rule[-6pt]{0pt}{19pt}}
& SU(2)$_L$ & SU(2)$_R$ & U(1)$_R$ & SO(8) & Sp($N$) \\
\hline \rule[-6pt]{0pt}{19pt}
& ~~ $A_\mu$ ~~ & $\mb1$ & $\mb1$ & 0 & $\mb1$ 
& $\mb{\mbit{N}(2\mbit{N}+1)}$ \\
\cline{2-7} $\scrV_{\rm adj}$ \rule[-6pt]{0pt}{19pt}
& $\lambda_\a$ & $\mb1$ & $\mb2$ & $+1$ & $\mb1$ 
& $\mb{\mbit{N}(2\mbit{N}+1)}$ \\
\cline{2-7} \rule[-6pt]{0pt}{19pt}
& $\v$ & $\mb1$ & $\mb1$ & $+2$ & $\mb1$ 
& $\mb{\mbit{N}(2\mbit{N}+1)}$ \\
\hline \rule[-6pt]{0pt}{19pt}
& $\phi$ & $\mb2$ & $\mb2$ & 0 & $\mb1$ 
& $\mb{\mbit{N}(2\mbit{N}-1)}$ \\
\cline{2-7} \raisebox{4pt}{$\scrH_{\rm a.s.}$} \rule[-6pt]{0pt}{19pt}
& $\psi_\a$ & $\mb2$ & $\mb1$ & $-1$ & $\mb1$ 
& $\mb{\mbit{N}(2\mbit{N}-1)}$ \\ 
\hline \rule[-6pt]{0pt}{19pt}
& $q$ & $\mb1$ & $\mb2$ & 0 & $\mb{8_{\rm v}}$ 
& $\mb{2\mbit{N}}$ \\
\cline{2-7} \raisebox{4pt}{$\scrH_{\rm f}$} \rule[-6pt]{0pt}{19pt} 
& $\eta_\a$ & $\mb1$ & $\mb1$ & $-1$ & $\mb{8_{\rm v}}$ 
& $\mb{2\mbit{N}}$ \\
\hline
\end{tabular}
\end{center}
\caption{{\small The SU(2)$_L\times$SU(2)$_R\times$U(1)$_R\times$SO(8) 
and Sp($N$) quantum  numbers of the elementary fields in the $\calN$=2
SYM theory.}}
\label{tqn}
\end{table}

The combination of $\calN$=2 supersymmetry and the above global
symmetries completely determines the form of the action of the theory,
which is given explicitly in section \ref{willines}.

AdS/CFT constructions for different $\calN$=2 theories have also been
proposed. In particular a U($N$) $\calN$=2 SYM theory with
hypermultiplets in the fundamental representation has been considered
in \cite{Karch:2002sh}. The set-up involves $N_f$ D7-branes wrapping
AdS$_5 \times S^3$ in the AdS$_5 \times S^5$ background obtained as
near horizon geometry of a stack of $N$ D3-branes. The resulting dual
gauge theory has U($N$) gauge group and the field content of $\calN$=4
SYM plus $N_f$ hypermultiplets in the fundamental representation. In
\cite{Karch:2002sh} it was argued that this configuration of D3- and
D7-branes is stable in spite of the fact that the $S^3$ contained in
$S^5$ wrapped by the D7-branes is contractible. At the level of
supergravity this follows from the observation that the scalar mode
corresponding to the D7-branes slipping off the $S^3$ although
`tachionic' does not violate the Breitenlohner--Freedman bound. At
leading order in $\al$ our analysis of the Higgs branch can be
repeated for this U($N$) theory essentially without
modifications. However in the following we shall be interested in
including $\al$ corrections in the discussion and it is not clear
whether the above stability argument would still apply at higher
orders. This is because unlike the configuration dual to the Sp($N$)
SYM theory, the D3-D7 system without the orientifold plane cannot
arise as a consistent string background. Moreover the Sp($N$) gauge
theory discussed above is finite for any $N$ whereas the U($N$) theory
with $N_f$ hypermultiplets has a $\b$-function (for the 't Hooft
coupling) which vanishes in the strict $N\to\infty$ limit, but is
positive for any finite $N$. We hope to generalise our discussion of
the Higgs branch to other models in the future.

\section{The Higgs branch}
\label{higgbr}

The Higgs branch of ${\cal N}=2$ theories consists of the vacua
for which the scalar components of the ${\cal N}=2$ fundamental
hypermultiplets get expectation values. In the context of the
above D3-D7-O7 system, vacuum expectation values for the
fundamental hypermultiplets correspond to D3-branes which are
``dissolved'' in the D7-branes. Dissolved D3-branes can be viewed
as instantons inside the SO(8) world-volume gauge theory of the
D7-branes \cite{Douglas:1995bn}. This is due to the Wess-Zumino
term in the D7-brane action,
\begin{align}
\mu_7 \int \, C_{\rm pb}^{(4)}\wedge {\rm tr}( F \wedge F) \, ,
\end{align}
where $F$ is the world volume field strength on the D7-branes, and
$C_{\rm pb}^{(4)}$ is the pull-back of the Ramond-Ramond four-form which
couples to D3-brane charge. A non-zero instanton number $K$ associated
with field strengths in the $4,5,6$ and $7$ directions corresponds to
$K$ D3-branes extended in the $0,1,2$ and $3$ directions and bound to
the D7-branes. It is well known that $F$- and $D$-flatness conditions
describing the Higgs branch can be mapped
\cite{Witten:1995gx,Douglas:1996uz} to the ADHM constraints
\cite{Atiyah:ri} determining the moduli space of instantons.

We will implement this general construction in the context of the
AdS/CFT correspondence in order to provide a holographic description
of the Higgs branch in the large $N$ limit of the above $\calN$=2
gauge theory. More precisely we will take as a starting point a stack
of $N=N^\pp+M$ (with $M\ll N^\pp$) D3-branes in the orientifold
geometry and we will consider the portion of the Higgs branch
corresponding to dissolving $M$ D3-branes in the world volume of the
D7-branes. In this way we can assume that the near horizon geometry of
the remaining $N^\pp$ D3-branes is not modified~\footnote{The background
we are considering, although not maximally supersymmetric, may be an
exact solution in the full string theory (see also \cite{swy}), in
which case the construction we are considering should not require the
above approximation.} and remains AdS$_5 \times S^5/Z_2$. The
resulting AdS/CFT construction describes a part of the Higgs branch in
which a Sp($N^\pp$) subgroup of the original Sp($N$) gauge group is
unbroken.

We will argue that the usual instantons, namely SO(8) field strengths
with support in the $X_\perp$ directions which are anti-self-dual with
respect to the flat metric $\dr X_\perp^2$, remain solutions of the
D7-brane equations of motion in the near horizon limit where the D7
geometry becomes (\ref{mmetric}). Such configurations provide a 
holographic description of (part of) the Higgs branch
of the boundary field theory.

The $\calN$=2 SYM theory we are studying has a rich structure of
global symmetries as discussed in the previous section. It is
interesting to analyse the pattern of global symmetry breaking in the
portion of the Higgs branch captured by our AdS construction. As
already remarked this corresponds to points where a residual
Sp($N^\pp$) gauge invariance is preserved. The moduli
identifying points on the Higgs branch correspond to the moduli of
the SO(8) $M$-instanton configuration on the
D7-brane world volume. Analysing the ADHM construction of instantons
for SO($n$) gauge theories \cite{Dorey:2002ik} we can characterise the
corresponding points on the Higgs branch of the dual theory. In
particular it is possible to identify the specific instanton
configurations associated with field theory vacua in which different
subgroups of the global symmetry are broken.

The logic here is the following. A generic instanton configuration
breaks some of the isometries of AdS$_5\times S^5/Z_2$ background as
well as the SO(8) gauge symmetry. In the dual boundary $\calN$=2 SYM
theory this corresponds to giving vev's to scalars that break the same
part of the conformal and global symmetries. By matching patterns of symmetry
breaking on the two sides one can determine a precise correspondence
between moduli of instantons and points on the Higgs branch. Clearly
in the presence of vev's for the scalar fields conformal invariance is
broken. This effect corresponds to the fact that the instantons we
consider break the SO(2,4) symmetry of the AdS$_5$ space, \eg via the
moduli associated with instanton locations in the radial
direction. Among the moduli on the Higgs branch there are vev's for
the scalars in the fundamental of Sp($N$) giving rise to non-vanishing
values for Sp($N^\pp$) invariant operators of the type
$\scrQ^{mn}=\widetilde q^m q^n$ (where gauge indices have been
suppressed). The quantities $\scrQ^{mn}$ are in the
$\mb1\oplus\mb{28}\oplus\mb{35}$ of SO(8). We can thus distinguish
points on the Higgs branch where the global SO(8) is unbroken,
corresponding to $\scrQ$'s in the singlet, or broken by vev's in
either the $\mb{28}$ or the $\mb{35}$. The vev's $\la q^n\ra$ for the
fundamental scalars can be identified with the instanton moduli,
$w^n$, associated with global SO(8) gauge orientations and the
$\scrQ^{mn}$'s correspond to combinations of such moduli.  
In the one-instanton sector the SO(8) singlet modulus corresponds to 
the combination $w^nw_n$ which is related to the instanton size
$\rho$. The (relative) positions of the $M$ instantons in the D7
world-volume theory generically break not only the SO(2,4) symmetry of
AdS$_5$, but also the SO(4) symmetry on $S^3$ and in particular its
SU(2)$_L$ subgroup. Since only fields belonging to the hypermultiplet
in the anti-symmetric second rank representation of Sp($N$) are
charged with respect to SU(2)$_L$ (see table \ref{tqn}) we conclude
that the AdS configuration we are considering captures regions of the
phase space in which vev's for the anti-symmetric hypermultiplet are
also turned on. More precisely, because of the unbroken Sp($N^\pp$), we
can have non-vanishing vev's for composite operators of the form
$\scrS= \widetilde\phi\phi$, where again gauge indices have not
been indicated, but it is understood that the operator is Sp($N^\pp$)
invariant. In this sense the phase of the theory that we are
describing can be considered as a generalised Higgs branch in that it
is characterised not only by fundamental hypermultiplet vev's, but by
non-zero vev's for the hypermultiplets in the anti-symmetric
representation as well. Analogously a generic instanton configuration
breaks the SU(2)$_R$ part of the R-symmetry. Notice however that the
SO(8) instantons cannot break the symmetry under SO(2) rotations in
the $(X^8,X^9)$ directions. This agrees with the fact that the vev's
of combinations of hypermultiplet scalars are not charged under
U(1)$_R$.

A more detailed analysis of the relevant multi-instanton ADHM
construction and of how it allows to precisely identify different
points on Higgs branch and characterise the corresponding pattern of
symmetry breaking is beyond our purposes here. We shall however return
to some of these aspects in section \ref{willines}, where we shall
consider non-local Wilson-line operators and show that their
expectation values coincide with the vev's parametrising the Higgs
branch.

\subsection{Instantons from D7 equations of motion at leading order\\ 
in $\alpha'$}

The D7-brane action is $S_{\rm D7} = S_{\rm DBI} + S_{\rm WZ}$ where 
the terms which involve only the field-strengths are
\begin{align}
\label{dbig} 
S_{\rm DBI} &= \frac{1}{(2\pi)^7 \gs\alpha'^4} \int \dr^8 x\,
\sqrt{-g} \,(2\pi\alpha')^2 \,\frac{1}{4}\,{\rm tr} (F_{ab}F^{ab})
\, + \cdots \\
S_{\rm WZ} &= \frac{1}{(2\pi)^7 \gs\alpha'^4} \int\, \sum_q C^{(q)}
\wedge\, {\rm tr}\, ( \er^{2\pi\alpha' F}) \, .
\end{align}
The ``$\cdots$'' in (\ref{dbig}) represents terms of higher order in
$\alpha'$. In the AdS background, the $\alpha'$ expansion is
effectively converted into an expansion in $1/\sqrt{\lambda}$, where
$\lambda = 4\pi\gs N$ is the 't Hooft coupling of the dual
gauge theory. The self-dual five-form field strength in AdS$_5 \times
S^5$ corresponds to a Ramond-Ramond four-form
\begin{align} 
C^{(4)}_{0123} = \frac{U^4}{L^4} \, .
\end{align}
Taking the AdS$_5 \times S^3$ D7-brane embedding (\ref{mmetric})
and turning on field strengths with support in the
$X_{\perp}$ directions, $X_4,\ldots,X_7$, one has
\begin{align}
S_{\rm DBI} = \frac{N}{(2\pi)^4\lambda L^4} \int \dr^4
x_\parallel \, \dr^4 X_\perp \, \ups^4 \,
\frac{1}{2}\,{\rm tr}(F_{mn}F_{mn}) + \cdots \label{dbi} \\
S_{\rm WZ} = \frac{N}{(2\pi)^4\lambda L^4} \int \dr^4
x_\parallel\, \dr^4 X_\perp\, \ups^4 \,
\frac{1}{4}\,\veps_{mnrs} {\rm tr}(F_{mn}F_{rs})  \, ,
\label{wesz}
\end{align} 
where roman indices indicate the directions $m=4,5,6,7$ and $\ups$ was
defined after equation (\ref{mmetric}).  Summing (\ref{dbi}) and
(\ref{wesz}) gives
\begin{align}
\label{cactn} 
S_{\rm D7} = \frac{N}{(2\pi)^4\lambda L^4}\int \dr^4 x_\parallel
\,\dr^4 X_\perp\, \ups^4 \,{\rm tr} (F^+)^2 \, ,
\end{align} 
where $F^+_{mn} \equiv \half\left(F_{mn} +
\frac{1}{2}\veps_{mnrs}F_{rs}\right)$.  Therefore, to at least
leading order, a field strength satisfying $F^+ =0$ gives also a
solution for the D7-branes embedded on AdS$_5 \times S^3$. Such a
configuration corresponds to an ordinary flat space instanton
solution. This is because, despite the curved background, $F_{mn}$ is
anti-self-dual with respect to a {\it flat} metric in the
$X^{4,5,6,7}$ directions and moreover with our choice of variables the
range of these coordinates is $(-\infty,+\infty)$.  Note that the
D7-brane action is the same (\ie vanishing) for all values of the
instanton number. This is expected, since the action of the string
backgrounds dual to Higgs branch vacua should be the same for all
such vacua.

\section{Higher order terms}
\label{hoterms}

The correspondence between the moduli space of Yang--Mills instantons
and the Higgs branch of ${\cal N} =2$ theories
\cite{Douglas:1995bn,Witten:1995gx,Douglas:1996uz} suggests that
connections which are anti self-dual with respect to a flat metric
should be exact solutions of the non-abelian D7-brane action in the
AdS background, even when all $1/\sqrt{\lambda}$ (or $\alpha'$)
corrections are taken into account.

Little is known about the exact form of the non-abelian D7-brane
action. The terms that are relevant for our analysis are those
involving powers of the gauge field strength and fields which have a
non vanishing background value in AdS$_5\times S^5/Z_2$, namely the
curvature, $\calR$, and \RR five-form, $\calF_5$. We are interested in
verifying that the anti self-dual configurations solving the field
equations at leading order remain solutions after including the $\al$
corrections. For this purpose we can consider separately the terms
involving only gauge fields and those involving couplings to $\calR$
and $\calF_5$, since the $F=F^-$ solution must be unaltered regardless
of the magnitude of $F_{mn}$ with respect to the background values of
$\calR$ and $\calF_5$.

The first corrections from terms involving only powers of the gauge
field strength are couplings of the form $F^4$ arising at order
$\al^2$, since the cubic terms of order $\al$ are known to
vanish. These $F^4$ terms are known exactly
\cite{Gross:1986iv,Tseytlin:1986ti,Tseytlin:1997cs,Bergshoeff:2001dc}.
Little is known about curvature couplings and essentially nothing is
known about the couplings to the \RR five-form. Some of the  ${\cal
R}^2 F^2$ are known \cite{Wijnholt:2003pw}, based on T-duality
arguments and a comparison with the $\sqrt{-g}{\cal R}^2$ terms
computed in \cite{Bachas:1999um,Fotopoulos:2001pt}.  The term in the
action of the form $\ups^4 F_{mn}{}^*F_{mn}$ is exactly given
by the CP-odd Wess-Zumino term (see \ref{wesz}), which is un-modified
by higher order terms in the $\alpha'$ expansion. For connections with
$F^+ = 0$ to solve the equations of motion, the CP-even terms
quadratic in $F$ must be of the form $\ups^4 F_{mn}F_{mn}$, with
exactly the same coefficient. This term is already present with the
correct coefficient in the absence of curvature and five-form
couplings to $F^2$. Thus the sum of the additional couplings of $F^2$ to
$\calR$ and $\calF_5$ must vanish when the curvature and five-form are
set to their AdS background values. The non-trivial order $\alpha'^2$
term in the AdS background is then just the $F^4$ term, ${\cal
L}_{F^4}$. To preserve the $F^+ =0$ solutions to this order requires
$\delta {\cal L}_{F^4}|_{F^+ =0} =0$. We will verify that this is the
case in the next section, and then turn our attention to the curvature
and five-form couplings.

\subsection{Terms of higher order in the field strength}
\label{higherf}

We first analyse higher order terms in the $\alpha'$ expansion of the
D7-brane action which only involve powers of the field
strength. Curvature and five-form couplings to $F^2$ in the AdS
background will be discussed in the next subsection.  There are no
cubic terms at order $\al$ and therefore the next to leading terms in
the effective action are of the form $F^4$
\cite{Gross:1986iv,Tseytlin:1986ti,bers,Koerber:2001uu}. More
precisely,
\ba
\hsp{-0.5}S_{F^4} &\!=\!& -\frac{1}{(2\pi)^5 \gs \alpha'^2}
\int \, \sqrt{-g}\: {\rm tr}\left[\alpha'^2\left(\frac{1}{24}
F_{AB}F^{BC}F_{CD}F^{DA} + \frac{1}{12}F_{AB}F^{BC}F^{DA}F_{CD}
\right.\right.\nonumber \\
&& \left.\left. \hsp{2}-\frac{1}{48}F_{AB}F^{BA}F_{CD}F^{DC}
-\frac{1}{96}F_{AB}F_{CD}F^{BA}F^{DC}\right)\right]\, ,
\label{ffour} 
\ea
where capital Roman indices indicate the real coordinates
$0,1,\ldots,7$. Taking the induced metric to be (\ref{mmetric}),
turning on field strengths only in the $X_\perp$ directions, and
including the Wess-Zumino term, the D7-brane action to order
$\alpha'^2$ can be written as 
\ba
&& S_{\rm D7} = \frac{N}{(2\pi)^4\lambda L^4}
\int\dr^4x_\parallel \dr^4X_\perp\, {\rm tr}\left[\ups^4 \,
F^+_{mn}F^+_{mn} \rule{0pt}{15pt} \right. \nn \\ 
&& \left. \hsp{3} - \frac{L^4}{12\lambda} \, \ups^8 \left(
2 F^+_{mn}F^+_{mn}F^-_{rs}F^-_{rs} +
F^+_{mn}F^-_{rs}F^+_{mn}F^-_{rs} \right)\right]\, ,
\label{f4terms}
\ea
where lower case Roman indices indicate the directions $4,5,6$ and
$7$, and $F^\pm_{mn}\equiv \half\left(F_{mn}\pm
\half\veps_{mnrs}F_{rs}\right)$. All terms (\ref{f4terms}) involve
two factors of $F^+$, therefore the associated equations of motion are
proportional to $F^+$ and thus are manifestly solved by anti-self-dual
connections.

The instanton solutions $F^+ =0$ can be viewed as stable
holomorphic bundles in two complex dimensions. In fact, a
requirement which has been used to constrain the form of the
non-abelian DBI action (modulo curvature and $p$-form couplings) is
that there should be BPS solutions which are holomorphic vector
bundles satisfying a {\it deformed} stability condition
\cite{Koerber:2001uu,Koerber:2002zb}, which reduces
to the usual condition in the $\alpha'\rightarrow 0$ limit. This
approach has recently been used to propose the form of the 
$O(\alpha'^4)$ terms \cite{Koerber:2002zb}, while checks of this
approach have been made to order $\alpha'^3$
\cite{Collinucci:2002ac}. In terms of complex coordinates,
$x^\alpha$, a holomorphic bundle satisfies
\begin{align}
F_{\alpha\beta} = F_{\bar\alpha\bar\beta} = 0 \, ,
\end{align}
while a stable holomorphic bundle also satisfies, in flat space,
\begin{align} 
\sum_\alpha F_{\alpha\bar\alpha}=0\, .
\end{align}
To order $\alpha'^2$, the equations of motion of a D-brane in a
flat background are solved by holomorphic bundles satisfying the
deformed stability condition
\begin{align}\label{defstab}
F_{\alpha\bar\alpha} -
\alpha'^2\frac{1}{6}\left(F_{\alpha\bar\beta} F_{\beta\bar\gamma}
F_{\gamma\bar\alpha} + F_{\alpha\bar\beta}F_{\gamma\bar\alpha}
F_{\beta\bar\gamma}\right) = 0\, .
\end{align}
In four real dimensions, with complex coordinates $x^1 + i x^2$
and $x^3+ix^4$, the holomorphic condition is $F_{13} = F_{24}$
and $F_{23} = -F_{14}$, while the stability condition is $F_{12}
= -F_{34}$. The deformed stability condition (\ref{defstab}) can
be written as
\ba
&& iF^+_{12} + \alpha'^2 \frac{i}{12}\left[{F_{12}^-}^2 F_{12}^+ +
F_{12}^- F_{12}^+ F_{12}^- + F_{12}^+ {F_{12}^-}^2 + {F_{12}^+}^3
\right. \nn \\
&& \hsp{2.2} \left. + F_{12}^+({F_{13}^-}^2 + {F_{14}^-}^2) + 
F_{13}^-\{F_{13}^-,F_{12}^+\} + F_{14}^-\{F_{14}^-,F_{12}^+\}
\rule{0pt}{14pt}\right] = 0 \, ,
\ea
which is clearly satisfied when $F^+ =0$. Note that, despite the
deformation of the stability condition, connections with $F^+ =0$
are solutions to at least order $\alpha'^2$ and presumably to all
orders in $\alpha'$.  As we have discussed and verified to order
$\alpha'^2$, the $F^+ =0$ solutions should also be exact in the
AdS background.

\subsection{Curvature and five-form couplings}

We now wish to consider relevant couplings of the D7-brane field 
strength to the 
bulk fields.  In the AdS background, the couplings which could modify the
field equations for the field strength on the D7-branes
involve the curvature tensors $\calR$, five-form $\calF_5$ and
their derivatives.  These include terms quadratic in the D7-brane 
field strength which are of the schematic form 
\be
\calL \sim D^r\calR^mD^s\calF_5^nF^2 \, , 
\qquad m,n,r,s=1,2,\ldots \, , 
\label{genrf5coupling}
\ee
arising at order $(\al)^{m+\half(n+r+s)}$. Terms of this type of
order $\al$ are expected to vanish. The absence of terms of this order
involving couplings of scalars to the curvature in the D-brane
effective action was verified in \cite{Fotopoulos:2002wy}, where terms
of the form $\calR X^2$, with $X$ a scalar field, were computed. As
noted in the previous section, the couplings of the form
(\ref{genrf5coupling}) must sum to zero when the
five-form and curvature tensors are set to their AdS values.  As we
will see shortly, this is not true of the {\it known} couplings of
this type, which are at order $\alpha'^2$.

Among the known couplings of D-branes to curvature are those
contained in the Wess-Zumino term \cite{Green:1996dd},
\begin{align}
\label{droof}
S_{\rm WZ} = \mu_p\int C\wedge{\rm tr}\left(\er^{2\pi\alpha' F}\right) 
\wedge \left(\frac{\hat{\cal A}(4\pi^2\alpha'R_T)}{{\hat{\cal A}
(4\pi^2\alpha'R_N)}}\right)^{1/2} \, ,
\end{align}
where $\hat{\cal A}$ is the ``A-roof genus'', which has an expansion
in terms of even powers of the curvature two-form. In (\ref{droof})
$R_T$ and $R_N$ are the tangent and normal bundle components of the
curvature respectively. The Wess-Zumino couplings of an orientifold
plane have a form similar to (\ref{droof}), but with $F$ set to zero
and the genus polynomials, $\hat\calA(R_T)$ and $\hat\calA(R_N)$,
replaced by the Hirzebruch polynomials, $\hat\calL(R_T/4)$ and
$\hat\calL(R_N/4)$, respectively \cite{Dasgupta:1997cd,mss,bste}. The
O-plane charge, $\mu_p^\pp$, is related to the D-brane charge, $\mu_p$,
by $\mu_p^\pp=-2^{p-4}\mu_p$.  In the following we shall only be
interested in vertices involving the field strength and therefore we
shall not consider the O-plane couplings.  In the AdS background, the
only non-zero Ramond-Ramond form is $C^{(4)}$, so that there are no
couplings of field strengths to curvature in the Wess-Zumino component
of the D7-brane action.

The $\calR^2$ terms in the D-brane effective action fixed by
two-graviton disk amplitudes were first obtained in
\cite{Bachas:1999um}. These terms are
\ba
&& S = \mu_p\int \sqrt{-g}\left[1-\frac{1}{24}
\frac{(4\pi^2\alpha')^2}{32\pi^2} \left((R_T)_{\a\b\g\d}
(R_T)^{\alpha\beta\gamma\delta} \right. \right. \nn \\
&& \hsp{2.5} \left.\left. - 2(R_T)_{\alpha\beta} (R_T)^{\alpha\beta} 
- (R_N)_{ab\alpha\beta}(R_N)^{ab\alpha\beta}+ 2\bar
R_{ab}\bar R^{ab}\right) \rule{0pt}{14pt} \right] \, ,
\label{curv1}
\ea
where the various curvature tensors appearing here are defined in
the appendix. For the special case of an embedding with vanishing
second fundamental form, the tensors
$(R_T)_{\alpha\beta\gamma\delta}$ and $(R_N)_{ab\alpha\beta}$ are
pull-backs of the bulk Riemann tensor to the tangent and normal
bundle, indicated by Greek and Roman indices respectively (we
emphasize that this is a change of notation from the previous
sections). The tensors $(R_T)_{\alpha\beta}$ and $\bar R_{ab}$
are not pull-backs of the bulk Ricci tensor, but are obtained from
contractions of tangent indices in the pull-backs of the Riemann
tensor. Specifically, for vanishing second fundamental form,
\begin{align}
\bar R_{ab} \equiv g^{\alpha\beta}R_{\alpha ab \beta},\qquad (
R_T)_{\alpha\beta} \equiv g^{\lambda\mu}
R_{\lambda\alpha\mu\beta}\, ,
\end{align} 
where $g_{\alpha\beta}$ is the induced metric on the D-brane. For the
AdS$_5 \times S^3$ embedding in AdS$_5 \times S^5$, the second
fundamental form vanishes (see the appendix).

Consistency of the couplings (\ref{curv1}) with T-duality implies
the existence of $\calR^2 F^2$ terms of the form
\cite{Wijnholt:2003pw}
\ba 
&& \hsp{-0.5} S =\mu_p (2\pi\alpha')^2 \int
\sqrt{-g}\; \frac{1}{4}{\rm tr}(F_{\alpha\beta}F^{\alpha\beta})\,
\left[1-\frac{1}{24}\frac{(4\pi^2\alpha')^2}{32\pi^2}
\left((R_T)_{\alpha\beta\gamma\delta}(R_T)^{\alpha\beta\gamma\delta}
\right.\right. \nn \\
&& \hsp{3} \left.\left.- 2 (R_T)_{\alpha\beta} (R_T)^{\alpha\beta} 
- (R_N)_{ab\alpha\beta}(R_N)^{ab\alpha\beta}+ 2\bar
R_{ab}\bar R^{ab}\right) \rule{0pt}{14pt} \right]\, . 
\label{curve2}
\ea
For D7-branes wrapping AdS$_5 \times S^3$ inside AdS$_5 \times S^5$, 
the curvature couplings in (\ref{curve2}) are non-vanishing
\be
(R_T)_{\alpha\beta\gamma\delta}(R_T)^{\alpha\beta\gamma\delta} - 
2(\hat R_T)_{\alpha\beta} (\hat R_T)^{\alpha\beta} -
(R_N)_{ab\alpha\beta}(R_N)^{ab\alpha\beta}+ 2\bar R_{ab}\bar
R^{ab} = -\frac{6}{25} L^2 \, ,
\label{nonzero}
\ee
where $L^2 = \sqrt{\lambda}\al$ is the square of the AdS$_5$ (or
$S^5$) curvature scalar. In light of (\ref{nonzero}), there must be
extra couplings to the background at order $\alpha'^2$, besides those
in (\ref{curve2}). There are many other possible terms of this order
which could be non-zero, but in order to determine them one would need
to compute the relevant string amplitudes in the AdS background.
There may be couplings involving mixed contractions between curvature
tensors and world-volume field strengths, which cannot be determined
from the known $R^2$ couplings \cite{Wijnholt:2003pw}). Furthermore
couplings of D-branes to pull-backs of the bulk Ricci tensor are also
not fixed by the disk amplitudes computed in \cite{Bachas:1999um}. The
bulk Ricci-tensor vanishes via the leading order equations of motion
in an expansion about a background without flux. However, both the
Ramond-Ramond five-form flux and Ricci tensor are non-vanishing in the
AdS background. Finally the contributions of $\calF_5$ couplings must
be taken into account. No couplings of the \RR five-form to world
volume gauge fields have been computed explicitly. The set of possible
terms in involving $\calF_5$  can be restricted and in principle
determined using supersymmetry arguments \cite{grst}.

The correspondence between the Higgs branch and Yang--Mills instantons
requires that the sum of all the above terms vanish for the AdS$_5
\times S^3$ embedding in AdS$_5 \times S^5$ and therefore provides
constraints order by order in $\al$ on the D-brane effective action.

\section{Non-renormalization of the Higgs Branch metric}
\label{metrnonren}

Thus far we have shown (to order $\alpha'^2$) that connections
with $F^+ =0$ solve the equations of motion of D7-branes wrapping
AdS$_5 \times S^3$, provided that certain constraints on the
curvature and five-form couplings are satisfied. To complete the
story, we should also show that the moduli-space approximation
\cite{Manton:1981mp} for
the dynamics of slowly moving instantons on a D7-brane in this
background reproduces the correct metric on the Higgs branch. This
metric is known to be equivalent to the metric describing the
dynamics of slowly varying instantons in eight-dimensional flat
space super Yang--Mills theory (see \cite{Dorey:2002ik} for a
review). To order $\alpha'^2$, we will find the correct metric,
assuming the cancellation of curvature and five-form couplings to
$F^2$ in the AdS background.

The gauge field part of the leading order action for a D7-brane
with the induced metric (\ref{mmetric}) is given by
\begin{align}\label{bigact}
S = \frac{N}{(2\pi)^4L^4\lambda} \int \dr^4x_\parallel
\dr^4X_\perp \, {\rm tr}\!\left( \fr{2\ups^4}  
\, F_{\mu\nu}F_{\mu\nu} + F_{m\mu}F_{m\mu} + \ups^4
F^+_{mn}F^+_{mn}\right) + \cdots\, .
\end{align}
Greek indices now indicate the $x_\parallel$ directions $0,1,2,3$
while Roman indices indicate the $X_\perp$ directions $4,5,6,7$.
To this order, the equations of motion are
\begin{align}
\label{eqnsmot}
D_m\left(\ups^4 F_{mn}^+\right) 
+ D_\mu F_{\mu n} = 0 \, , \qquad 
D_m F_{m\nu} + \fr{\ups^4} D_\mu F_{\mu\nu} = 0 \, .
\end{align}
Let us write the instanton solutions as $ A_m = \bar A_m (X_\perp,
{\cal M}^i)$, $A_{\mu}=0$, where ${\cal M}^i$ are the
instanton moduli. These solutions are exact provided the $\calM^i$'s 
have no dependence on $x_\parallel$. Slowly varying instantons may
be studied by considering the gauge field configuration
\begin{align}
\label{bacg}
A_m = \bar A_m(X_\perp, {\cal M}^i(x_\parallel)), \qquad 
A_\mu = \Omega_i\partial_\mu {\cal M}^i \, ,
\end{align}
where $\Omega$ is Lie algebra valued. Choosing $\Omega_i$ such
that the equations of motion (\ref{eqnsmot}) are satisfied to
linear order in derivatives with respect to $x_\parallel$ requires
\begin{align}
\label{provided}
D_m \left(\frac{\partial \bar A_m}{\partial {\cal M}^i} -
D_m\Omega_i\right) =0 \, .
\end{align}
Inserting (\ref{bacg}) into the (\ref{bigact}) gives the following
effective action for the collective coordinates to quadratic
order in $\partial_\mu {\cal M}^i$
\ba
S &\!\!=\!\!&\frac{N}{(2\pi)^4L^4\lambda} \int \dr^4x_\parallel 
\dr^4 X_\perp \,\frac{1}{4}\, \delta_i \bar A_m \delta_j\bar A_m 
\partial_{\mu}{\cal M}^i \partial_{\mu} {\cal M}^j \nn \\
&& = -\frac{N}{(2\pi)^4L^4\lambda} \int \dr^4x_\parallel
\, \half\, G_{ij}(\calM) \, \del_\mu\calM^i \del_\mu\calM^j \, ,
\label{modspappr}
\ea
where 
\begin{align} \delta_i \bar A_m \equiv \frac{\partial \bar A_m}
{\partial {\cal M}^i} - D_m\Omega_i 
\end{align} 
and we have introduced the metric on the moduli space,
$G_{ij}(\calM)$, which,  at leading order in the $1/\sqrt{\lambda}$
expansion, is therefore
\begin{align}\label{HM}
G_{ij}(\calM) = -\half \int \dr^4 X_\perp \, \delta_i 
\bar A_m(X_\perp,\calM) \, \delta_j \bar A_m(X_\perp,\calM) \, .
\end{align}
This is equivalent to the metric for Yang--Mills instantons in flat
space, which captures the exact metric on the Higgs branch, so we do
not expect corrections at higher orders.

The metric on the Higgs branch of ${\cal N}=2$ gauge theories is
known to be given exactly by the tree level result
\cite{Argyres:1996eh}. This result is obtained by pulling back the
flat metric associated with the tree level hypermultiplet kinetic
terms to the space determined by the $F$- and $D$-flatness
constraints, subject to a quotient by the gauge symmetry. Therefore
the only dependence on the gauge coupling should be an overall
$1/\gy^2 = N/\lambda$ in front of the moduli space action. The tree
level result should be the leading (and only non-trivial) term in the
strong coupling expansion which is obtained from the AdS/CFT duality.

The next to leading term in the strong coupling expansion of the
metric is obtained from the order $\alpha'^2$ terms in the DBI
action. To this order,
\ba
&& \hsp{-1} \partial_\mu{\cal M}^i \partial_\mu \calM^j 
G_{ij}(\calM) = - \int \dr^4
X_\perp\,\frac{1}{2}{\rm tr}(F_{\mu m} F_{\mu m}) \nn \\
&& \hsp{2.5} + \frac{L^4}{\lambda}\int \dr^4
X_\perp\,\frac{\ups^4}{6}{\rm tr} \left[F_{s\mu}F_{\mu n}
\left(\{F_{nr},F_{rs}\} -\frac{1}{2}\delta_{sn}F_{tu}F_{ut}\right)
\right. \nn \\ 
&& \hsp{2.5} + \frac{1}{2}\left. \left(F_{\mu n}F_{nr}F_{s\mu}F_{rs} 
+ F_{\mu n}F_{rs}F_{s\mu}F_{nr} - \frac{1}{2}F_{\mu n}
F_{rs}F_{n\mu}F_{sr}\right)\right]\, .
\label{horder}
\ea
Rewriting the field strengths with all components in the $X_\perp$
directions in terms of self-dual and anti-self-dual components,
(\ref{horder}) becomes
\ba
\label{additional}
&& \hsp{-1}\partial_\mu{\cal M}^i \partial_\mu {\calM}^j G_{ij}
(\calM) = - \int \dr^4 X_\perp\,\frac{1}{2}\, 
{\rm tr}(F_{\mu m} F_{\mu m}) \\
&& \hsp{0.5} + \frac{L^4}{\lambda}\int\, \dr^4 x_\perp
\,\frac{\ups^4}{12}\,{\rm tr} \left[F_{s\mu}F_{\mu n}
\left(\{F^+_{nr},F^-_{rs}\}+\{F^-_{nr},F^+_{rs}\}
\right)\right. \nn \\
&& \hsp{0.5} + \frac{1}{2}\left. \left(F_{\mu n}F^+_{nr}F_{s\mu}
F^-_{rs}+ F_{\mu n}F^-_{nr}F_{s\mu}F^+_{rs} + F_{\mu n}F^+_{rs}F_{s\mu}
F^-_{nr}+F_{\mu n}F^-_{rs}F_{s\mu}F^+_{nr} \right)\right]\, . \nn
\ea
To order $\alpha'^2$, the equations of motion are still solved by
(\ref{bacg}) to linear order in derivatives with respect to
$x_\parallel$ provided that (\ref{provided}) is satisfied.
Furthermore the additional contributions to the metric arising from
the order $1/\lambda^2$ terms in (\ref{additional}) vanish when
$F^+=0$. The non-renormalization of the metric on the Higgs branch
implies the absence of corrections to (\ref{HM}) at any order.

\section{Non-renormalization of chiral Wilson lines}
\label{willines}

In this section we shall present a proof a new non-renormalization
property for eight-supercharge Yang-Mills theories in dimension $d\le
4$. Specifically, we will show that certain straight BPS Wilson
lines with scalar components of hypermultiplets at the endpoints have
length independent expectation values. These expectation values are
the same as those of local operators parameterizing the Higgs
branch. The non-renormalization theorem we will find is very similar
to one conjectured in \cite{Zarembo:2002an} and demonstrated in
\cite{Guralnik:2003di} for a class of BPS Wilson loops in maximally
supersymmetric Yang-Mills theories. Our result is also closely
related to the non-renormalization of the metric on the Higgs branch,
for reasons that will become clear shortly.

We will obtain our result by making a particularly useful sub-group of
the full supersymmetry manifest. In the context of ${\cal N} =2$ gauge
theories in four-dimensions, we will write the action using ${\cal
N}=2$ {\it three dimensional}  superspace~\footnote{The idea of
writing the action for a supersymmetric theory in terms of a lower
dimensional superspace has been discussed and applied in a variety of
other situations
\cite{Marcus:1983wb,Arkani-Hamed:2001tb,Erdmenger:2002ex,Constable:2002xt,
Constable:2002vt,Erdmenger:2003kn,Dijkgraaf:2003xk}.}. The ${\calN}=2$
four-dimensional supersymmetry algebra generated by $Q_{i\alpha}$,
$\bar Q^i_{\dot\alpha}$ is given (in the absence of central charge) by
\begin{align}
& \{Q_{i\alpha},\bar Q^j_{\dot\beta}\} =
2\sigma^\mu_{\alpha\dot\beta}P_\mu\delta^j_i \nonumber \\
& \{Q_{i\alpha},Q_{j\beta}\} = \{\bar
Q^i_{\dot\alpha},\bar Q^j_{\dot\beta}\} =0
\end{align}
where $i=1,2$ is the SU(2) R-symmetry index and $\alpha$=1,2 is a
spinor index. This algebra contains two copies of an $\calN=2$, $d=3$ 
supersymmetry algebra. Defining
\begin{align}
& Q_\alpha \equiv \frac{1}{2}(Q_{1\alpha}+\bar Q^1_{\dot\alpha}) +
\frac{i}{2}(Q_{2\alpha}+\bar Q^2_{\dot\alpha})\nonumber\\
& \hat Q_\alpha \equiv \frac{i}{2}(Q_{1\alpha}-\bar
Q^1_{\dot\alpha}) - \frac{1}{2}(Q_{2\alpha}-\bar
Q^2_{\dot\alpha})\, ,
\end{align}
The ${\cal N}=2$, $d=4$ algebra becomes
\begin{align}
& \{Q_\alpha,\bar Q_\beta\} = 2\sigma^M_{\alpha\beta}P_M \, ,
\qquad M= 0,1,3  \nonumber \\
& \{\hat Q_\alpha,\bar {\hat Q}_\beta\} =
2\sigma^M_{\alpha\beta}P_M \\
& \{Q_\alpha,\bar {\hat Q}_\beta\} = \{\hat Q_\alpha, \bar
Q_\beta\} = -2i\sigma^2_{\alpha\beta}P_2\, ,
\end{align}
with all other commutators vanishing.  The supercharges $Q_\alpha$ and
$\bar Q_\beta$ (or $\hat Q_\alpha$ and $\bar {\hat Q}_\beta$)
generate an ${\cal N}=2$, $d=3$ supersymmetry.  In the next section we
re-write the ${\cal N}=2$ theory  in a superspace which makes an
${\cal N}=2$, $d=3$ supersymmetry and not an ${\cal N}=1$, $d=4$
supersymmetry manifest.  A remarkable feature of this superspace is
that a class of straight Wilson lines can be written as bottom
components of chiral superfields.  The constraints on the chiral ring
can then be used to make exact statements about expectations values of
these Wilson lines.

\subsection{${\cal N}=2$, $d=4$ SYM in ${\cal N}=2$, $d=3$ superspace}

We wish to write the action for the four-dimensional ${\cal N}=2$
Sp($N$) SYM theory that we have considered in the previous sections in
terms of three-dimensional ${\cal N} =2$ superspace. It is
straightforward to generalize the analysis in this section to other
${\cal N}=2$ theories. Similar non-renormalisation properties for
straight Wilson-line operators can be proven for a large class of
eight supercharge Yang-Mills theories with hypermultiplets.

The action of the $\calN$=2 Sp($N$) SYM theory can be written in 
terms of conventional four-dimensional $\calN$=1 superfields as
\ba 
S &\!=\!& \fr{\gy^2} \int \dr^4x \left\{ \fr{16}\left[\int
\dr^2\theta \, \tr\left( \calW^\a \calW_\a \right) + \, {\rm h.c.} 
\right] + \int \dr^2\theta \dr^2\bar\theta \, \tr \left( \er^{-V}
\Phi^\dagger \er^V \Phi \right) \right. \nn \\
&& \hsp{0.4} + \int \dr^2\theta \dr^2\bar\theta \left[
\Psi^\dagger_r \left(\er^V\right)^i \left(T^i_{\rm
a.s.}\right)^{rs}\Psi_s + \widetilde\Psi^\dagger_r
\left(\er^{-V}\right)^i \left(T^i_{\rm a.s.}
\right)^{rs}\widetilde\Psi_s \right] \nn \\
&& \hsp{0.4} + \int \dr^2\theta \dr^2\bar\theta \left[
(\calQ^{9-n})^\dagger_a\left(\er^V\right)^i \left( T^i_{\rm f} 
\right)^{ab} \calQ^n_b + \widetilde \calQ^{n\,\dagger}_a 
\left(\er^{-V}\right)^i \left(T^i_{\rm f} \right)^{ab} 
\widetilde \calQ^{9-n}_b \right] \nn \\
&& \hsp{0.4} \left. + \left[ \int \dr^2\theta \left(
\widetilde \Psi_r \Phi^i \left(T^i_{\rm a.s.}\right)^{rs} 
\Psi_s + \widetilde \calQ^{9-n}_a \Phi^i \left(T^i_{\rm f}
\right)^{ab} \calQ^n_b\right) + \, {\rm h.c.} \right] \right\} \, , 
\label{N1action} 
\ea
where we have denoted by $V$ and $\Phi$ the $\calN$=1 vector and
chiral multiplets in the adjoint of Sp($N$) which form the $\calN$=2
vector multiplet, by $\Psi$ and $\widetilde\Psi$ the two $\calN$=1
chiral multiplets in the second rank anti-symmetric hypermultiplet and
by $\calQ$ and $\widetilde \calQ$ the chiral multiplets forming the
hypermultiplets in the fundamental. The traces are over matrices in
the fundamental used to represent the fields in the $\calN$=2 vector
multiplet. The $T^i_{\rm a.s.}$'s, $i=1,\ldots,N(2N+1)$ are generators
in the anti-symmetric and the indices $r$ and $s$ run from 1 to
$N(2N-1)$, whereas the $T^i_{\rm f}$'s in the fundamental have indices
$a$ and $b$ running from 1 to $2N$. The index $n=1,\ldots,4$ is used
to label the fundamental of SO(8). The hypermultiplets $\calH^u_{\rm
f}$, $u=1,\ldots,8$, are decomposed into $\calN$=1 chiral multiplets
$\calQ^1,\ldots,\calQ^4$ and $\widetilde \calQ^5,\dots,\widetilde
\calQ^8$ and similarly  their conjugates $\calH^{u\,\dagger}_{\rm f}$
into $\widetilde \calQ^{1\,\dagger},\ldots,\widetilde
\calQ^{4\,\dagger}$ and
$\calQ^{5\,\dagger},\ldots,\calQ^{8\,\dagger}$. The field strength
superfield $\calW_\a$ is defined as usual as
\be 
\calW_\a = -\fr{4} \bar D \bar D \er^{-V} D_\a \er^V \, . 
\label{superfs} 
\ee 
In (\ref{N1action}) a gauge fixing term has not been indicated
explicitly.

Note that the dimensional reduction of ${\cal N}=1$, $d=4$ superspace
to three dimensions gives ${\cal N}=2$, $d=3$ superspace.  As a
starting point for obtaining the above action in an ${\cal N}=2$,
$d=3$ superspace, one can first dimensionally reduce (\ref{N1action})
on a circle (say the $x^3$ direction), giving a three dimensional
eight supercharge theory written in ${\cal N}=2$, $d=3$ superspace
(which is thus manifestly invariant under four real
supersymmetries). One can then re-introduce the non-zero modes in the
$x^3$ direction, while keeping the dimension of the superspace
fixed. The three-dimensional ${\cal N} =2$ superfields used to
describe a four-dimensional ${\cal N}=2$ theory have the general form
$F(x^0,x^1,x^2,\theta,\bar\theta| X^3)$, where the superspace is
spanned by $x^0,x^1,x^2,\theta,\bar\theta$ and $X^3$ is to be viewed
as a continuous label.  An ${\cal N}=2$, $d=4$ vector multiplet
corresponds to a continuous set of ${\cal N}=2$, $d=3$ vector
superfields $V(X^3)$ together with a continuous set of  ${\cal N}=2$,
$d=3$ chiral superfields $\Phi(X^3)$. An ${\cal N}=2$, $d=4$
hypermultiplet corresponds to a continuous set of doublets of ${\cal
N}=2$, $d=3$ chiral multiplets, $\calQ(X^3)$ and $\widetilde
\calQ(X^3)$.

Aside from the continuous index, the superfield content of an ${\cal
N}=2$, $d=4$ theory in ${\cal N}=2$, $d=3$ superspace is basically the
same as in ${\cal N}=1$, $d=4$ superspace.  However the component
fields are distributed amongst the superfields differently.  For the
Sp($N$) theory we have been considering,  the requisite ${\cal N}=2$,
$d=3$ superfields  are again a vector superfield $V$, adjoint chiral
$\Phi$, anti-symmetric chirals $\Psi$ and $\widetilde\Psi$, and
fundamental chirals $\calQ^m$ and $\widetilde \calQ^{9-m}$. The ${\cal
N}=2$, $d=4$ vector multiplet contains the gauge connections
$A_{0,1,2,3}$ and adjoint Hermitian scalars $X^1, X^2$, which are
distributed amongst the ${\cal N}=2$, $d=3$ superfields $V$ and $\Phi$
as follows
\begin{align}
V&\rightarrow A_{0,1,2} \, , \; X^1 \,, \nn \\
\Phi&\rightarrow A_3 \, , \; X^2 \, . \label{4dphi}
\end{align}
The bottom component of $\Phi$ is $A_3 + i X^2$.  The remarkable
fact that a gauge connection belongs to a chiral superfield will
be used to obtain exact results for expectation values of Wilson
lines. Henceforward,  all superfields we write are 
in ${\cal N}=2, d=3$ superspace.   

An example of an ${\cal N}=2$, $d=4$ Yang-Mills action in ${\cal N
}=2$, $d=3$ superspace was written in \cite{Erdmenger:2002ex}. For the
case which we consider the structure is essentially the same.  For our
purposes the most  important term in the ${\cal N}=2$, $d=3$
superspace representation of the action is the superpotential, which
is given by
\ba  
{\rm W} &\!\!=\!\!& \int \dr X^3\,
\dr^3 x\, \dr^2 \theta\, \left[ \widetilde \calQ^{9-n}_a 
\left( i\d^{ab}\partial_{X^3} - \Phi^i \left(T^i_{\rm f}\right)^{ab}
\right) \calQ^n_b \right. \nn \\
&& \left. \hsp{1} + \widetilde\Psi_r \left(i\d^{rs}\del_{X^3} 
- \Phi^i \left(T^i_{\rm a.s.}\right)^{rs} \right) \Psi_s + 
{\rm tr}\left({\cal W}^{\alpha}\calW_\alpha\right) \right]\,. 
\label{suplag}
\ea 
Although the details of the K\"ahler potential will not be be important
in the subsequent discussion, we record it below for completeness
\ba
{\rm K} &\!\!=\!\!& \fr{\gy^2} \int \dr X^3\,
 \dr^3x \left\{   \int \dr^2\theta \dr^2\bar\theta \, \tr \left( 
\er^{-V} \Phi'^\dagger \er^V \Phi' \right) \right. \nn \\
&& \hsp{0.4} + \int \dr^2\theta \dr^2\bar\theta \left[
\Psi^\dagger_r \left(\er^V\right)^i \left(T^i_{\rm
a.s.}\right)^{rs}\Psi_s + \widetilde\Psi^\dagger_r
\left(\er^{-V}\right)^i \left(T^i_{\rm a.s.}
\right)^{rs}\widetilde\Psi_s \right] \nn \\
&& \hsp{0.4} + \left. \int \dr^2\theta \dr^2\bar\theta 
\left[(\calQ^{9-n})^\dagger_a \left(\er^V\right)^i 
\left( T^i_{\rm f} \right)^{ab} \calQ^n_b +
\widetilde \calQ^{n\,\dagger}_a \left(\er^{-V}\right)^i \left(
T^i_{\rm f} \right)^{ab} \widetilde \calQ^{9-n}_b \right] 
\right\} \, , 
\label{kah} 
\ea 
where 
\begin{align}
\Phi' \equiv \Phi + e^{-V}(i\partial_{X^3} - \bar\Phi)e^{V}
\end{align}
Note that the kinetic terms for fields arising from ${\cal N}=2$,
$d=4$ hypermultiplets involving derivatives with respect to $X^3$
arise from the superpotential in ${\cal N}=2$, $d=3$ superspace rather
than the K\"ahler potential.  This fact, together with
four-dimensional Lorentz invariance, can be used to show the absence
of radiative corrections to the metric on the Higgs branch. We will
use the fact that a kinetic term is contained in the superpotential to
show that the expectation values of a class of straight BPS Wilson
lines are the same as the expectation values of local operators
parameterizing the Higgs branch.

\subsection{Chiral Wilson lines}

One can define Wilson lines in the four-dimensional ${\cal N}=2$
theory which are chiral with respect to ${\cal N}=2$, $d=3$
supersymmetry.  This is possible because  the bottom component of the
adjoint ${\cal N}=2$, $d=3$ chiral superfield contains a gauge
connection, $\Phi = A_3 + i X^2 + \cdots$.  Under gauge
transformations parameterized by  ${\cal N}=2$, $d=3$ superfields
$\Lambda(X^3)$,  $\Phi$ transforms as
\begin{align}
\Phi \rightarrow \er^{i\Lambda} \Phi \er^{-i\Lambda} -
i\,\er^{i\Lambda} \frac{\partial}{\partial x^3}\er^{-i\Lambda}\, .
\end{align}
Thus a gauge invariant chiral superfield whose components are Wilson 
lines is given by
\begin{align}
\label{eqns}
\scrW_{\rm f}^{mn}(X^3) \equiv \widetilde \calQ^{9-m}_a(0) \,{\cal P}
\exp\left(i\int_0^{X^3} \dr X'^3 \, \Phi \right)^{ab} \calQ^n_b(X^3)\, .
\end{align}
Note that this Wilson line is straight, extending only in the $X^3$
direction transverse to the superspace. The chiral structure is lost
if one considers Wilson lines which are not straight.

$\scrW^{mn}(X^3)$ in equation (\ref{eqns}) 
decomposes into $\mb1\oplus\mb{28}\oplus\mb{35}$ with
respect to the SO(8) global symmetry. 
The expectation value of the bottom component, 
\bdm
W_{\rm f}^{mn}=\scrW_{\rm f}^{mn}\big |_{\theta=\bar\theta=0} \, , 
\edm
satisfies 
\ba
\partial_{X^3}\langle W_{\rm f}^{mn}(X^3) \rangle &\!\!=\!\!& 
\langle \left. \widetilde \calQ^{9-m}_a(0) {\cal P} 
\exp\left(i\int_0^{X^3} \dr X'^3 \,\Phi \right)^{ac}\left(
\partial_{X^3} + i \Phi(X^3)\right)^{cb} \calQ_b^n(X^3) 
\right|_{\theta=\bar\theta=0}\rangle \nn \\
&\!\!=\!\!& -i\langle \left. \widetilde \calQ_a^{9-m}(0) {\cal P} 
\exp\left(i\int_0^{X^3} \dr X'^3 \, \Phi\right)^{ac}
\frac{\delta {\rm W}}{\delta \widetilde \calQ^{9-n}_c(X^3)}
\right|_{\theta=\bar\theta =0} \rangle \, ,
\label{nonreno}
\ea
where ${\rm W}$ is the superpotential (\ref{suplag}). 

Consider an infinitesimal variation of the form
\begin{align}
\label{infvar}
\widetilde \calQ_c^{9-n}(X^3) \rightarrow \widetilde \calQ_c^{9-n}(X^3) 
+ \epsilon f^n_c(X^3)
\end{align}
where $f_c(X^3)$ is a functional of the chiral superfields. 
Specifically, we choose 
\begin{align}
\label{choice} 
f^n_c(X^3) =\widetilde \calQ_b^{9-n}(0) {\cal P} \exp\left(i\int_0^{X^3} 
\dr X'^3 \, \Phi(X'^3)\right)^{bc} \, .
\end{align} 
The classical equation derived from this variation is
\begin{align}
\label{cleqn}
\bar D^2\left[ f_a^m \left(\er^{-V}\right)^{ab} 
\widetilde \calQ^{n\dagger}_b \right] = f^m_a \frac{\delta {\rm W}}{\delta
\widetilde \calQ_a^{9-n}} \, .
\end{align}
Classical equations of this form are often modified quantum
mechanically, giving rise to what is known as a generalized
Konishi anomaly \cite{Cachazo:2003yc}. In appendix
\ref{Anomaly} we show that there is no Konishi anomaly in this case.

In a supersymmetric vacuum $\langle \bar D^2(\cdots)|_{\theta
=\bar\theta =0} \rangle = 0$. Hence equation (\ref{cleqn})
implies $\langle f^m_a \frac{\delta {\rm W}}{\delta \widetilde 
\calQ^{9-n}_a} \rangle = 0$, so that (\ref{nonreno}) becomes
\begin{align}
\partial_{X^3}\la W_{\rm f}^{mn}(X^3) \ra = 0 \, . 
\label{lengthind}
\end{align} 
Therefore the expectation value of these Wilson lines is the same as
that of a local operator parameterizing the Higgs branch, 
\begin{align}
\label{tensionless}
\langle W_{\rm f}^{mn}(L) \rangle = \langle W_{\rm f}^{mn}(0) \rangle
= \langle \widetilde q^m_a q^n_a 
\rangle \, ,
\end{align} 
where $\widetilde q$ and $q$ are the bottom components of $\widetilde
\calQ$ and $\calQ$ respectively. 

As already observed, the operators $W_{\rm f}^{mn}$ decompose with
respect to the SO(8) global symmetry into
\ba
&& W_{{\rm f};\,\mb1} = W_{\rm f}^{mm} \nn \\
&& W^{[mn]}_{{\rm f};\,\mb{28}} = \half \left( W_{\rm f}^{mn}-
W_{\rm f}^{nm} \right) \label{vevsdecomp} \\ 
&& W^{\{mn\}}_{{\rm f};\,\mb{35}} = \half \left( W_{\rm f}^{mn} 
+ W_{\rm f}^{nm} \right) - \fr{8} \,\d^{mn} W_{\rm f}^{kk} \, . \nn
\ea
At different points on the Higgs branch some or all of the
combinations (\ref{vevsdecomp}) can be non-zero. As a result the SO(8)
global symmetry is unbroken if only $W_{{\rm f};\,\mb1}$ is
non-vanishing, or broken if either $W^{[mn]}_{{\rm f};\,\mb{28}}$ or
$W^{\{mn\}}_{{\rm f};\,\mb{35}}$  have a non-vanishing expectation
value.

The same analysis can be repeated without any modification for
straight Wilson lines constructed from the hypermultiplets in the
anti-symmetric representation. In this case one defines
\be
\scrW_{\rm a.s.}(X^3) \equiv \widetilde \Psi_r(0) {\cal P}
\exp\left(i\int_0^{X^3} \dr X'^3\,\Phi \right)^{rs} \Psi_s(X^3)
\label{anstisymW}
\ee
and 
\be
W_{\rm a.s.} = \scrW_{\rm a.s.}\big |_{\theta=\bar\theta=0} \, .
\label{botWas}
\ee
By the same steps described above one can show that 
\be
W_{\rm a.s.}(X^3) = W_{\rm a.s.}(0) = \la \widetilde\phi_r\phi_r\ra 
\, , \label{tensionlessas}
\ee
 where $\widetilde \phi$ and $\phi$ are the bottom components of 
$\widetilde \Psi$ and $\Psi$ respectively.

Analogous arguments can be applied in other $\calN$=2 SYM theories,
\eg in the case of the U($N$) theory with $N_f$ hypermultiplets in
fundamental considered in the AdS/CFT context in \cite{Karch:2002sh}.
In these theories one can define similar BPS Wilson-line operators
whose expectation values are independent of the length and
parameterize the  Higgs branch.  Furthermore the arguments leading to
equation (\ref{lengthind})  are unmodified upon dimensionally reducing
in directions belonging to the ${\cal N} =2$, $d=3$ superspace. Thus
(\ref{lengthind}) also applies to eight-supercharge Yang--Mills
theories in $3,2$ and $1$ dimension~\footnote{Notice that in 1
dimension, the superspace one  uses to show (\ref{lengthind}) is
actually zero dimensional.}. However the situation is different for
five dimensional ${\cal N}=1$ theories,  for which there is a
non-trivial generalized Konishi anomaly (see appendix \ref{Anomaly}).
\begin{figure}[!ht]
\begin{center}
\includegraphics[width=0.6\textwidth]{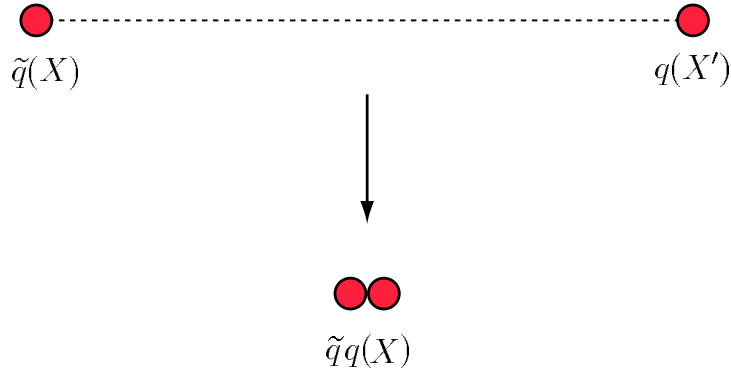}
\caption{{\small Chiral Wilson lines with the same expectation values.
The circles indicate scalar components of fundamental hypermultiplets. 
Wilson lines with anti-symmetric hypermultiplets at the end points
have the same property.}}
\end{center}
\end{figure}

In the case of five-dimensional theories one uses a four-dimensional
$\calN$=1 superspace and it is possible to derive a classical equation
like (\ref{cleqn}).  However in this case there is a quantum anomaly, 
as is shown in appendix \ref{Anomaly}.

\section{Conclusions and future directions}
\label{concl}

In this article we have presented a proposal for the dual AdS
description of the Higgs branch of a finite four-dimensional ${\cal
N}=2$ Sp($N$) gauge theory with one hypermultiplet in the second rank
anti-symmetric representation and four hypermultiplets in  the
fundamental representation. The theory has a SO(8) flavour symmetry
which corresponds to a bulk gauge symmetry associated with the
presence of D7-branes. There is a one to one map between instantons in
the eight-dimensional SO(8) super Yang-Mills theory on the
world-volume of the D7-branes  and the Higgs branch of the ${\cal
N}=2$ Sp($N$) gauge theory. Although the AdS description involves
D7-branes in a curved background with Ramond-Ramond flux,  the
equations of motion at leading order in $\alpha'$ admit solutions
corresponding to field strengths which are anti-self-dual with respect
to a flat four dimensional metric, \ie ordinary SO(8)
instantons. Furthermore the metric describing the dynamics of slowly
varying instantons on the D7-brane world-volume reproduces the correct
metric on the Higgs branch.

To order $\alpha'^2$, we have checked that terms in the D7-action
which only involve field strengths are of the right form to preserve
the anti-self-dual solutions and give the correct metric on the Higgs
branch. Little is known about bulk couplings to the D7-brane beyond
leading order in $\alpha'$. The correspondence with the Higgs branch
of the dual gauge theory in the set-up that we have considered,
together with known non-renormalisation theorems for $\calN$=2 SYM
theories, leads to constraints on the higher-derivative terms in the
D-brane effective action in the AdS background. The absence of
corrections to the metric on the Higgs branch requires that the
anti-self-dual solutions of the D7-brane field equations are not
modified by the inclusion of $\al$ corrections. The known couplings at
order $\al^2$ do not satisfy this requirement and therefore
additional, yet to be determined, couplings to the curvature and/or
\RR five-form must be present.

The $\alpha'$ corrections to bulk couplings of D-branes are very
important in the context of brane-world physics (see \cite{afrey} for
example) and for obtaining the strong coupling expansion of gauge
theories with fields in the fundamental representation using AdS/CFT
duality. The supergravity realization of the Higgs branch
non-renormalization is not sufficient to determine the bulk-brane
couplings, but does provide an important constraint. It would be
interesting to consider the supergravity description of the Higgs
branch of other eight supercharge Yang-Mills theories, which should
yield additional insights into this issue.

We have also presented a new non-renormalization theorem for straight
BPS Wilson lines in eight-supercharge Yang--Mills theories. This
non-renormalization theorem is closely related to the physics of the
Higgs branch. We have shown that open Wilson lines with hypermultiplet
insertions at the end points have length independent expectation
values. This result was shown using a lower-dimensional superspace
which makes a particularly useful subgroup of the full extended
supersymmetry manifest. In this superspace, the Wilson lines are
bottom components of chiral superfields and the non-renormalisation
result follows from equations satisfied by the corresponding chiral
ring. It would be interesting to study whether similar superspace
methods can be used to obtain other non-renormalization theorems.

We expect that the non-renormalisation properties of the Wilson-line
operators be preserved when instanton effects are included. Adapting
the methods developed in \cite{bgk} to the case of $\calN$=2 theories
and using the BPS nature of these Wilson lines, the analysis of
fermion zero-modes in the instanton background should allow to prove
the absence of instanton corrections to their expectation values.

Another interesting open question is whether the expectation values of
BPS open Wilson lines in the conformal Sp($N$) ${\cal N}=2$ theory can
be computed using the AdS/CFT duality, in a way analogous to the AdS
computation of the expectation values of closed Wilson loops in ${\cal
N}=4$ super Yang-Mills theory \cite{maldaloops,soojong1,drukker}. One
might expect the open Wilson line expectation value to be captured by
a semiclassical approximation to an open string partition function,
where the open string has endpoints on D7-branes. The result should
depend on the point on the Higgs branch through holonomies $\exp(i\int
A)$ involving the gauge field of an SO(8) instanton background on the
D7-branes.

\vsp{0.7}
\ndt
{\large {\bf Acknowledgments}} 

\vsp{0.3}
\ndt
We wish to thank M. Bianchi, G. Cardoso, J. Erdmenger, M. Green,
A. Hanany, D. L\"{u}st, J. Maldacena, D. Mateos, H. Nastase,
K. Peeters, S. Ramgoolam, H. Schnitzer, W. Skiba, S. Stieberger and
M. Zamaklar for useful discussions.

\appendix

\section{Embedding geometry}

Here we very briefly summarize some of the facts about embedding
geometry relevant to this article. A more detailed discussion of
the geometry of embeddings may be found in \cite{Bachas:1999um}
and references therein. Consider a $d$-dimensional submanifold of a
D-dimensional space described by $Y^A(\zeta^\alpha)$, where $A =1,
\ldots, D$ and $ \alpha = 1,\ldots, d$. Pull-backs of bulk tensors
to the tangent bundle are defined by contractions with
$\partial_{\alpha}Y^A$, while pull backs to the normal bundle are
defined by contractions with $\xi^A_{a}$, where lower case roman
indices correspond to the normal bundle and
\begin{align}
\xi^A_{a}\xi^B_{b}G_{AB} =\delta_{ab}, \qquad
\xi^A_a\partial_\alpha Y^B G_{AB} =0\, .
\end{align}
Raising and lowering of indices in the tangent bundle is defined
with respect to the induced metric,
\begin{align} g_{\alpha\beta} =
\partial_\alpha Y^A \partial_\beta Y^B G_{AB}\, ,\end{align}
while indices in the normal bundle are raised or lowered with
respect to $\delta_{ab}$.

The second fundamental form is defined as the covariant derivative
of the tangent frame $\partial_\alpha Y^A$
\begin{align}
\Omega^A_{\alpha\beta} = \Omega^A_{\beta\alpha} \equiv
\partial_\alpha\partial_\beta Y^A -
(\Gamma_T)^\gamma_{\alpha\beta} \partial_\gamma Y^A +
\Gamma^A_{BC}\partial_\alpha Y^B \partial_\beta Y^C\, ,
\end{align}
where $\Gamma_T$ is the Levi-Civita connection constructed from
$g_{\alpha\beta} =
\partial_\alpha Y^A \partial_\beta Y^B G_{AB}$. In terms of the
second fundamental form and pull-backs of the bulk Riemann tensor,
the curvature tensors appearing in (\ref{curv1}) are defined as
follows;
\begin{align}
(R_T)_{\alpha\beta\gamma\delta} &\equiv
R_{\alpha\beta\gamma\delta} +
\delta_{ab}(\Omega^a_{\alpha\gamma}\Omega^b_{\beta\delta} -
\Omega^a_{\alpha\delta}\Omega^b_{\beta\gamma}) \\
(R_N)_{\alpha\beta}{}^{ab} &\equiv -R^{ab}{}_{\alpha\beta} +
g^{\gamma\delta}(\Omega^a_{\alpha\gamma}\Omega^b_{\beta\delta} -
\Omega^b_{\alpha\gamma}\Omega^a_{\beta\delta})\\
(\hat R_T)_{\alpha\beta} &\equiv (R_T)^\gamma{}_{\alpha\gamma\beta}\\
{\bar R}_{ab} &\equiv R^\alpha{}_{ab\alpha} +
g^{\alpha\gamma}g^{\beta\delta}\Omega_{a|\alpha\beta}
\Omega_{b|\gamma\delta}
\end{align}
The various curvature tensors appearing in (\ref{curv1}) are
\be 
(R_T)_{\alpha\beta\gamma\delta} = \left\{ \begin{array}{ll}
\displaystyle -\frac{L}{20}(g_{\alpha\gamma}g_{\beta\delta} -
g_{\alpha\delta}g_{\beta\gamma}) \rule{0pt}{19pt} 
& \qquad {\rm for ~ AdS_5 ~ indices} \\
\displaystyle \frac{L}{20}(g_{\alpha\gamma}g_{\beta\delta} -
g_{\alpha\delta}g_{\beta\gamma}) \rule{0pt}{21pt}
& \qquad {\rm for} ~ S^3 ~ {\rm indices} \nonumber \\
\displaystyle 0 \rule{0pt}{19pt} 
& \qquad{\rm for ~ mixed ~ indices} 
\end{array} \right.
\ee
\be
(R_T)_{\alpha\beta} = \left\{ \begin{array}{ll} 
\displaystyle -\frac{L}{5}g_{\alpha\beta} \rule{0pt}{19pt}
& \qquad {\rm for ~ AdS_5 ~ indices} \\
\displaystyle \frac{L}{10}g_{\alpha\beta} \rule{0pt}{19pt}
& \qquad {\rm for} ~ S^3 ~ {\rm indices} \\
\displaystyle 0 \rule{0pt}{19pt}
& \qquad{\rm for ~ mixed ~ indices} 
\end{array} \right.
\ee
\be
(R_N)_{ab\alpha\beta} = 0 \, ,
\ee
\be
\bar R_{ab} = -\frac{3}{20}L\delta_{ab} \, ,
\ee
where $L = (\lambda \al^2)^{1/4}$ is the scalar
curvature of $S^5$, or minus the scalar curvature of AdS$_5$.

\section{Variation of the functional measure}
\label{Anomaly}

In this appendix we investigate if there is a non-zero anomalous
modification of the Ward identities derived in (\ref{cleqn}).  We will
compute the variation of the functional measure under (\ref{infvar})
using methods discussed in \cite{Konishi:1985tu,Cachazo:2003yc}. Our
discussion will closely parallel that of \cite{Guralnik:2003di}.

To be somewhat more general, we consider an eight supercharge theory
in $d+1$ dimensions written in a $d$-dimensional four-supercharge
superspace,
\ba
S &\!\!=\!\!& \int \dr X\, \dr^d y\, \dr^2 \theta\, \left( 
\widetilde \calQ_m (i\partial_{X} - \Phi) \calQ_m + {\rm tr}\,
{\cal W}_{\alpha}{\cal W}^\alpha\right) \nonumber \\
&\!\!+\!\!\!& \int \dr X \dr^d y\, \dr^4 \theta\, 
\left(\calQ^\dagger_m \er^V \calQ_m + \widetilde \calQ_m \er^{-V} 
\widetilde \calQ^\dagger_m + {\rm tr} ({\Phi^\pp}^\dagger \er^V 
\Phi^\pp \er^{-V})\right) \, ,
\ea
where $V$ and $\Phi$ are the $d$-dimensional ${\cal N} =1$ vector and
scalar  multiplets and $\calQ$, $\widetilde \calQ$ are ${\cal N}=1$
chiral multiplets in any  representation of the gauge group and $m$ is
a flavour index.

Under the infinitesimal transformation
\begin{align}
\widetilde \calQ_n(X) \rightarrow \widetilde \calQ_n(X) + \epsilon
f_{n,X}[\calQ,\widetilde \calQ, \Phi, \calW_\alpha] \, ,
\end{align}
the Jacobian is
\begin{align}\label{jac}
J = 1 + {\rm tr}_c\, \epsilon \bar D^2 \frac{\delta
f_{m,X}}{\delta \widetilde \calQ_{m}(X)} = 1+ \int \dr Z_c\,\sum_i\,
\epsilon(Z) \langle Z, i| \bar D^2 \frac{\delta f^m}{\delta 
\widetilde \calQ^m}|Z, i\rangle \, ,
\end{align}
where $Z$ collectively denotes the superspace coordinates, $z=(\vec
y, \theta,\bar\theta)$, and the transverse coordinate $X$, $\dr Z_c$ is
the chiral measure $\dr^dy\, \dr^2\theta\, \dr X$, and we  have
defined a Hilbert space spanned by $|Z,i\rangle$, which are eigenstates
of the coordinate operator $Z$ and belong to the same 
representation of the gauge group under which the matter fields
transform 
\begin{align}
\hat Z |Z,i\rangle = Z |Z,i\rangle,\qquad \hat T^A |Z,i\rangle =
T^A_{ij} |Z,j\rangle\, .
\end{align}
such that
\begin{align}
\frac{\delta \widetilde \calQ_{mi}(Z^{\prime})}
{\delta \widetilde \calQ_{nj}(Z)} =
\delta_{mn}\,\delta_{ij}\,\delta(X - X') \,\delta^d(y-y^{\prime})
\,\delta^2(\theta - \theta^{\prime}) =\delta_{mn}\, \la Z,i|\bar
D^2|Z',j\ra \, .
\end{align}
The matrix 
\begin{align} 
\langle Z,i| \frac{\delta f_{m,X}}
{\delta \widetilde \calQ_m(X)} (-\fr{4}\bar D^2) |Z', i\rangle
\end{align}
is proportional to $\delta^2(\theta -\theta') = (\theta - \theta')^2$
and so has vanishing diagonal entries. Thus naively $J = 1$. However
this is not necessarily true upon regularizing the trace.

To obtain the Jacobian for the transformation $\widetilde \calQ_m\rightarrow
\widetilde \calQ_m + \epsilon f^m$, we need to compute a regularized version
of the diagonal matrix element,
\begin{align}\label{diag}
{\cal M}_{X, z, i} \equiv\, \la X, z, i| \bar D^2 \frac{\delta
f^m}{\delta \widetilde \calQ^m}|X, z, i\ra \, .
\end{align}
The regularization used in \cite{Konishi:1985tu} in the more
familiar context a four-dimensional ${\mathcal N}=1$ gauge theory
involves insertion of an operator $\exp(-\hat L/M^2)$, where
\begin{align}
\hat L \equiv -\frac{1}{16}\bar D^2 \er^{-V}D^2\er^V \, .
\end{align}
Note that this operator is gauge covariant and chiral. However,
the insertion of $\exp(-\hat L/M^2)$ will not suffice in our case,
since this only cuts off large momenta in directions belonging to
the superspace. The regularized version of (\ref{diag}) which we
will consider is
\begin{align}\label{reg}
{\cal M}_{X, z, i} \equiv\, \la X, z, i| \exp(-\hat{\cal
L}/M^2)\,\bar D^2 \, \frac{\delta f^m}{\delta \widetilde \calQ^m}
|X, z, i\ra \, ,
\end{align}
where 
\begin{align} \hat {\cal L}
= \hat L + \left(\frac{\partial}{\partial X} + i \Phi\right)^2 \, .
\end{align}
To evaluate (\ref{reg}), note that
\begin{align}
\label{hatl}
\hat L (\bar D^2\cdots) = (\partial_t^2 - {1/2}
{\calW}^{\alpha}D_{\alpha} + C\partial_t + F)(\bar D^2 \cdots) \, ,
\end{align}
where
\begin{align}
& C \equiv \frac{1}{2}\bar D_{\alpha}\er^{-V} \bar D_{\alpha} \er^V
\nonumber \\
& F \equiv \frac{1}{16}\bar D^2 \er^{-V}D^2\er^V \, .
\end{align}
We can write
\begin{align}
\er^{-{\cal L}/M^2} = \er^{-\nabla^2/M^2} \hat {\cal S}\, ,
\end{align} 
where $\nabla^2$ is the Laplacian in the space including all bosonic
coordinates, $\nabla^2 \equiv \partial_{X}^2 + \nabla_{y}^2$~. The
factor $\hat{\cal S} = 1 + \cdots$ must contain a term with two
$D_{\alpha}$ operators for $\exp(\hat{\cal L}/M^2)\bar D^2$ to give a
non-zero contribution to (\ref{reg}). To illustrate this property,
note that
\begin{align} 
\la z^{\prime}| D^2 \bar D^2|z\ra &= \delta^d(y-y') \,D^2
\bar D^2 \,\delta^2(\theta -\theta')\,\delta^2(\bar\theta -
\bar\theta') \nonumber \\ 
&= \delta^d(y-y') \,D^2 \bar D^2 (\theta-\theta^{\prime})^2
\,(\bar\theta - \bar\theta^{\prime})^2 = \delta^d(y-y')\, .
\end{align}
If the $D^2$ were removed, the diagonal matrix element would
vanish. Thus, (\ref{hatl}) implies that, in a large $M^2$
expansion, the leading non-zero contribution to (\ref{reg}) is
\be
\label{expr1}
\hsp{-0.5}{\cal M}_{X, z, i} = 
\la X, z, i|\exp(-\nabla^2/M^2) 
\frac{W_{\alpha}W^{\alpha}}{M^4}D^2\bar D^2|X', z', i'\ra\la X', z',
i'|\frac{\delta f^m}{\delta \widetilde \calQ^m} |X, z, i\ra + \cdots
\ee
where
\begin{align}
\la X', z', i'|\frac{\delta f^m}{\delta \widetilde \calQ^m} |X, z, i\ra
\equiv \frac{\delta f^m_{X', i'}}{\delta \widetilde \calQ^m_{X, i}}
\,\delta^d(\vec y' -\vec y)\,\delta^2(\theta
-\theta')\,\delta^2(\bar\theta' -\bar\theta) \, .
\end{align}
Equation (\ref{expr1}) can be evaluated by inserting the identity
$|k_X,\vec k_y,j\ra\la k_X,\vec k_y,j|$\,\, after $\exp(-\nabla^2/M^2)$,
where $|k_X,\vec k_y,j\ra$ is an eigenvector of the momentum
operators in the transverse direction $X$ and the bosonic part of
the superspace $z$. The result is
\ba
\hsp{-1}{\cal M}_{X, z,i} &\!\!=\!\!& \int \dr k_X\, \dr^d k_y\, 
\dr X'\, \dr^dy'\, \exp\left(-\frac{k_X^2 + \vec k_y^2}{M^2} + 
i k_X (X-X') + i\vec k_y \cdot(\vec y -\vec y')\right) \nonumber \\
&& \frac{1}{M^4}\,W_{\alpha}^D(X', z') \,W^{\alpha E}(X',z')\, 
\la i|\hat T^D \hat T^E| i'\ra \, \frac{\delta f^m_{X', i'}}
{\delta \widetilde \calQ^m_{X, i}}\, \delta^d(\vec y' -\vec y) \nn \\
&\!=\!& M^{d-4}\,W_{\alpha}^D(X, z) \,W^{\alpha E}(X,z)\, 
\la i|\hat T^D \hat T^E| i'\ra \frac{\delta f^m_{X, i'}}{\delta 
\widetilde \calQ^m_{X, i}} \, .
\ea
The anomaly vanishes if the superspace is less than
four-dimensional $(d<4)$, but it is non-trivial when $d=4$,
corresponding to a five-dimensional ${\cal N} =1$ theory. This
anomaly is crucial for the validity of Dijkgraff-Vafa conjectures
relating effective superpotentials to auxiliary matrix models (or
an auxiliary matrix quantum mechanics in this case)
\cite{Cachazo:2003yc,Dijkgraaf:2003xk,Bena:2003tf}. It does not
make sense to discuss $d>4$, since there is no four supercharge
superspace in more than four dimensions.



\begin{thebibliography}{99}

\bibitem{Maldacena:1997re}
J.~M.~Maldacena, ``The large N limit of superconformal field
theories and supergravity'', \atmp{2}{1998}{231} [\ijtp{38}{1999}{1113}]
[\hepth{9711200}].

\bibitem{Gubser:1998bc}
S.~S.~Gubser, I.~R.~Klebanov and A.~M.~Polyakov, ``Gauge theory
correlators from non-critical string theory'', 
\plb{428}{1998}{105} [\hepth{9802109}].

\bibitem{Witten:1998qj}
E.~Witten, ``Anti-de Sitter space and holography'',
\atmp{2}{1998}{253} 
[\hepth{9802150}].

\bibitem{Fayyazuddin:1998fb}
A.~Fayyazuddin and M.~Spalinski, ``Large N superconformal gauge
theories and supergravity orientifolds'', \npb{535}{1998}{219} 
[\hepth{9805096}].

\bibitem{Aharony:1998xz}
O.~Aharony, A.~Fayyazuddin and J.~M.~Maldacena, ``The large N
limit of N = 2,1 field theories from three-branes in F-theory'',
\jhep{07}{1998}{013} [\hepth{9806159}].

\bibitem{Gross:1986iv}
D.~J.~Gross and E.~Witten, ``Superstring Modifications Of
Einstein's Equations'', \npb{277}{1986}{1}.

\bibitem{bers}{E.~Bergshoeff, M.~Rakowski and E.~Sezgin,
``Higher Derivative Super-Yang-Mills Theories'', \plb{185}{1987}{371}.}

\bibitem{Tseytlin:1986ti}
A.~A.~Tseytlin, ``Vector Field Effective Action In The Open
Superstring Theory'', \npb{276}{1986}{391}
[Erratum-ibid. {\bf B291} (1987) 876].

\bibitem{Tseytlin:1997cs}
A.~A.~Tseytlin, ``On non-abelian generalisation of the Born-Infeld
action in string  theory'', \npb{501}{1997}{41}
[\hepth{9701125}].

\bibitem{Bergshoeff:2001dc}
E.~A.~Bergshoeff, A.~Bilal, M.~de Roo and A.~Sevrin,
``Supersymmetric non-abelian Born-Infeld revisited'', 
\jhep{07}{2001}{029} [\hepth{0105274}].

\bibitem{Koerber:2001uu}
P.~Koerber and A.~Sevrin, ``The non-Abelian Born-Infeld action
through order $\al^3$'', \jhep{0110}{2001}{003} [\hepth{0108169}].

\bibitem{Koerber:2002zb}
P.~Koerber and A.~Sevrin, ``The non-abelian D-brane effective
action through order $\al^4$'', \jhep{10}{2002}{046} 
[\hepth{0208044}].

\bibitem{Collinucci:2002ac}
A.~Collinucci, M.~De Roo and M.~G.~C.~Eenink, ``Supersymmetric
Yang--Mills theory at order $\al^3$'', \jhep{06}{2002}{024}
[\hepth{0205150}].

\bibitem{Douglas:1995bn}
M.~R.~Douglas, ``Branes within branes'', \hepth{9512077}.

\bibitem{Douglas:1996uz}
M.~R.~Douglas, ``Gauge Fields and D-branes'', 
\jgp{28}{1998}{255} [\hepth{9604198}].

\bibitem{Witten:1995gx}
E.~Witten, ``Small Instantons in String Theory'', \npb{460}{1996}{541} 
[\hepth{9511030}].

\bibitem{Dorey:2002ik}
N.~Dorey, T.~J.~Hollowood, V.~V.~Khoze and M.~P.~Mattis, ``The
calculus of many instantons'', \prep{371}{2002}{231}
[\hepth{0206063}].

\bibitem{gk}{A.~Giveon and D.~Kutasov, ``Brane dynamics and gauge
theory'', \rmp{71}{1999}{983} [\hepth{9802067}].}

\bibitem{Bachas:1999um}
C.~P.~Bachas, P.~Bain and M.~B.~Green, ``Curvature terms in
D-brane actions and their M-theory origin'', \jhep{05}{1999}{011}
[\hepth{9903210}].

\bibitem{Fotopoulos:2001pt}
A.~Fotopoulos, ``On $\al^2$ corrections to the D-brane action
for non-geodesic world-volume embeddings'', \jhep{09}{2001}{005}
[\hepth{0104146}].

\bibitem{Wijnholt:2003pw}
M.~Wijnholt, ``On curvature-squared corrections for D-brane
actions'', \hepth{0301029}.

\bibitem{grst}{M.~B.~Green and C.~Stahn,
``D3-branes on the Coulomb branch and instantons'', 
\jhep{09}{2003}{052} [\hepth{0308061}].}

\bibitem{Argyres:1996eh}
P.~C.~Argyres, M.~R.~Plesser and N.~Seiberg, ``The Moduli Space of
N=2 SUSY {QCD} and Duality in N=1 SUSY {QCD}'', 
\npb{471}{1996}{159} [\hepth{9603042}].

\bibitem{Guralnik:2003di}
Z.~Guralnik and B.~Kulik, ``Properties of chiral Wilson loops'',
\hepth{0309118}.

\bibitem{Dijkgraaf:2002fc}
R.~Dijkgraaf and C.~Vafa, ``Matrix models, topological strings,
and supersymmetric gauge theories'', \npb{644}{2002}{3}
[\hepth{0206255}].

\bibitem{Dijkgraaf:2002dh}
R.~Dijkgraaf and C.~Vafa, ``A perturbative window into
non-perturbative physics'', \hepth{0208048}.

\bibitem{Dijkgraaf:2003xk}
R.~Dijkgraaf and C.~Vafa, ``N = 1 supersymmetry, deconstruction,
and bosonic gauge theories'', \hepth{0302011}.

\bibitem{Cachazo:2003yc}
F.~Cachazo, N.~Seiberg and E.~Witten, ``Chiral Rings and Phases of
Supersymmetric Gauge Theories'', \jhep{04}{2003}{018}
[\hepth{0303207}].

\bibitem{Bena:2003tf}
I.~Bena and R.~Roiban, ``N = 1* in 5 dimensions: Dijkgraaf-Vafa
meets Polchinski-Strassler'', \jhep{11}{2003}{001} 
[\hepth{0308013}].

\bibitem{mg}{M.~Gutperle, ``Heterotic/type I duality, D-instantons and
a $\calN$ = 2 AdS/CFT correspondence'',
\prd{60}{1999}{126001} [\hepth{9905173}].}

\bibitem{gns}{E.~Gava, K.~S.~Narain and M.~H.~Sarmadi,
``Instantons in $\calN$=2 Sp($N$) superconformal gauge theories and
the AdS/CFT correspondence'', \npb{569}{2000}{183}
[\hepth{9908125}].}

\bibitem{th}{T.~J.~Hollowood,
``Instantons, finite $\calN$=2 Sp($N$) theories and the AdS/CFT
correspondence'', \jhep{11}{1999}{012}
[\hepth{9908201}].}

\bibitem{bgmnn}{D.~Berenstein, E.~Gava, 
J.~M.~Maldacena, K.~S.~Narain and H.~Nastase,
``Open strings on plane waves and their Yang-Mills duals'',
\hepth{0203249}.}

\bibitem{sen}{A.~Sen, ``F-theory and Orientifolds'',
\npb{475}{1996}{562} [\hepth{9605150}].}

\bibitem{swy}{H.~J.~Schnitzer and N.~Wyllard, ``An orientifold of 
AdS$_5\times T^{11}$ with D7-branes, the associated  $\alpha'^2$
corrections and their role in the dual $\calN$=1  
Sp(2$N$+2$M$)$\times$Sp(2$N$) gauge theory'',
\jhep{08}{2002}{012} [\hepth{0206071}].}

\bibitem{Karch:2002sh}
A.~Karch and E.~Katz, ``Adding flavor to AdS/CFT'',
\jhep{06}{2002}{043} [\hepth{0205236}].

\bibitem{Atiyah:ri}
M.~F.~Atiyah, N.~J.~Hitchin, V.~G.~Drinfeld and Y.~I.~Manin,
``Construction Of Instantons'', \pla{65}{1978}{185}.

\bibitem{Green:1996dd}
M.~B.~Green, J.~A.~Harvey and G.~W.~Moore, ``I-brane inflow and
anomalous couplings on D-branes'', \cqg{14}{1997}{47} 
[\hepth{9605033}].

\bibitem{Dasgupta:1997cd}
K.~Dasgupta, D.~P.~Jatkar and S.~Mukhi, ``Gravitational couplings
and Z(2) orientifolds'', \npb{523}{1998}{465} [\hepth{9707224}].

\bibitem{mss}
J.~F.~Morales, C.~A.~Scrucca and M.~Serone,
``Anomalous couplings for D-branes and O-planes'',
\npb{552}{1999}{291} [\hepth{9812071}].

\bibitem{bste}
B.~Stefanski, ``Gravitational couplings of D-branes and O-planes'', 
\npb{548}{1999}{275} [\hepth{9812088}].

\bibitem{Zarembo:2002an}
K.~Zarembo, ``Supersymmetric Wilson loops'', \npb{643}{2002}{157} 
[\hepth{0205160}].

\bibitem{Marcus:1983wb}
N.~Marcus, A.~Sagnotti and W.~Siegel, ``Ten-Dimensional
Supersymmetric Yang--Mills Theory In Terms Of Four-Dimensional
Superfields'', \npb{224}{1983}{159}.

\bibitem{Arkani-Hamed:2001tb}
N.~Arkani-Hamed, T.~Gregoire and J.~Wacker, ``Higher dimensional
supersymmetry in 4D superspace'', \jhep{03}{2002}{055}
[\hepth{0101233}].

\bibitem{Erdmenger:2002ex}
J.~Erdmenger, Z.~Guralnik and I.~Kirsch, ``Four-dimensional
superconformal theories with interacting boundaries or defects'',
\prd{66}{2002}{025020} [\hepth{0203020}].

\bibitem{Constable:2002xt}
N.~R.~Constable, J.~Erdmenger, Z.~Guralnik and I.~Kirsch,
``Intersecting D3-branes and holography'', \prd{68}{2003}{106007} 
[\hepth{0211222}].

\bibitem{Constable:2002vt}
N.~R.~Constable, J.~Erdmenger, Z.~Guralnik and I.~Kirsch,
``(De)constructing intersecting M5-branes'', \prd{67}{2003}{106005} 
[\hepth{0212136}].

\bibitem{Erdmenger:2003kn}
J.~Erdmenger, Z.~Guralnik, R.~Helling and I.~Kirsch, ``A
world-volume perspective on the recombination of intersecting
branes'', \hepth{0309043}.

\bibitem{Konishi:1985tu}
K.~I.~Konishi and K.~I.~Shizuya, ``Functional Integral Approach To
Chiral Anomalies In Supersymmetric Gauge Theories'', 
\nc{A90}{1985}{111}.

\bibitem{Fotopoulos:2002wy}
A.~Fotopoulos and A.~A.~Tseytlin, ``On gravitational couplings in
D-brane action'', \jhep{12}{2002}{001} [\hepth{0211101}].

\bibitem{bds}{T.~Banks, M.~R.~Douglas and N.~Seiberg,
``Probing F-theory with branes'', \plb{387}{1996}{278}
[\hepth{9605199}].}

\bibitem{Manton:1981mp}
N.~S.~Manton, ``A Remark On The Scattering Of BPS Monopoles'',
\plb{110}{1982}{54}.

\bibitem{maldaloops}
J.~M.~Maldacena, ``Wilson loops in large $N$ field theories'',
\prl{80}{1998}{4859} [\hepth{9803002}].

\bibitem{soojong1}
Soo-Jong Rey and Jung-Tay Yee, ``Macroscopic strings as heavy quarks:
Large-$N$ gauge theory and anti-de Sitter supergravity'',
\epjc{22}{2001}{379} [\hepth{9803001}].

\bibitem{drukker} N.~Drukker, D.~J.~Gross and H.~Ooguri, ``Wilson 
loops and minimal surfaces'', \prd{60}{1999}{125006}
[\hepth{9904191}].

\bibitem{bgk}{M.~Bianchi, M.B.~Green and S.~Kovacs,
``Instanton corrections to circular Wilson loops in $\scrN$=4
supersymmetric Yang--Mills'', \jhep{04}{2002}{040} [\hepth{0202003}];
``Instantons and BPS Wilson loops'', \hepth{0107028}.}

\bibitem {afrey}
A.~R.~Frey, ``String theoretic bounds on Lorentz-violating warped 
compactification'', \jhep{04}{2003}{012} [\hepth{0301189}].

\bibitem{polyakov}
A.~M.~Polyakov, ``Fine Structure Of Strings'', \npb{268}{1986}{406}.

\bibitem{kleinert} H.~Kleinert, ``The Membrane Properties Of Condensing
Strings'', \plb{174}{1986}{335}.


\end{thebibliography}
\end{document}